\documentclass[twocolumn]{aastex631} 

\newcommand{\Msun}{\ensuremath{\,\rm{M}_{\odot}}\xspace}

\newcommand{\yr}{\ensuremath{\,\rm{yr}}\xspace}

\newcommand{\Gpc}{\ensuremath{\,\rm{Gpc}}\xspace}

\newcommand{\MbhA}{\ensuremath{M_{\rm BH, a}}\xspace}
\newcommand{\MbhB}{\ensuremath{M_{\rm BH, b}}\xspace}
\newcommand{\dMsn}{\ensuremath{dM_{\rm SN}}\xspace}
\newcommand{\MZAMSI}{\ensuremath{M_{\rm ZAMS, a}}\xspace}
\newcommand{\MZAMSII}{\ensuremath{M_{\rm ZAMS, b}}\xspace}
\newcommand{\Mcore}{\ensuremath{M_{\rm core}}\xspace}
\newcommand{\McoreA}{\ensuremath{M_{\rm core,a}}\xspace}
\newcommand{\McoreB}{\ensuremath{M_{\rm core,b}}\xspace}

\newcommand{\Mmm}{\ensuremath{M_{\rm primary}}\xspace}
\newcommand{\Mless}{\ensuremath{M_{\rm secondary}}\xspace}
\newcommand{\minmbh}{\ensuremath{\min(M_{\rm primary})}\xspace}
\newcommand{\MtildeII}{\ensuremath{M_{\rm post, MT1}}\xspace}

\newcommand{\Rlocal}{\ensuremath{\mathcal{R}_{0.2}}\xspace}

\newcommand{\qzams}{\ensuremath{q_{\mathrm{ZAMS}} }\xspace}
\newcommand{\qcritI}{\ensuremath{q_{\mathrm{crit,1}} }\xspace}
\newcommand{\qcritII}{\ensuremath{q_{\mathrm{crit,2}} }\xspace}
\newcommand{\qpremtI}{\ensuremath{q_{\mathrm{pre MT, 1}} }\xspace}
\newcommand{\qpremtII}{\ensuremath{q_{\mathrm{pre MT,2}} }\xspace}
\newcommand{\qfinal}{\ensuremath{q_{\mathrm{final}} }\xspace}

\newcommand{\fcore}{\ensuremath{f_{\mathrm{core}} }\xspace}

\newcommand{\asn}{\ensuremath{a_{\mathrm{SN}} }\xspace}
\newcommand{\bsn}{\ensuremath{b_{\mathrm{SN}} }\xspace}
\newcommand{\betaacc}{\ensuremath{\beta_{\mathrm{acc}} }\xspace}
\newcommand{\zetaeff}{\ensuremath{\zeta_{\mathrm{eff}} }\xspace}

\newcommand{\fmix}{\ensuremath{f_{\mathrm{mix}}}\xspace}
\newcommand{\fdisk}{\ensuremath{f_{\mathrm{disk}}}\xspace}

\newcommand{\COMPAS}{{\tt COMPAS}\xspace}

\usepackage{xspace}
\usepackage{cancel}
\usepackage{amsmath}
\usepackage{hyperref}
\usepackage{lipsum}
\usepackage{amsmath}
\usepackage{graphicx} 
\graphicspath{{./}{figures/}} 

\shorttitle{stable mass transfer and the apparent NS--BH gap}
\shortauthors{van Son et al.}

\begin{document}

\title{No peaks without valleys: The stable mass transfer channel for gravitational-wave sources in light of the neutron star--black hole mass gap.}

\correspondingauthor{L.~van Son}
\email{lieke.van.son@cfa.harvard.edu}

\author[0000-0001-5484-4987]{L.~A.~C.~van~Son}
\affiliation{Center for Astrophysics $|$ Harvard $\&$ Smithsonian,
60 Garden St., Cambridge, MA 02138, USA}
\affiliation{Anton Pannekoek Institute for Astronomy, University of Amsterdam, Science Park 904, 1098XH Amsterdam, The Netherlands}
\affiliation{Max Planck Institute for Astrophysics, Karl-Schwarzschild-Strasse 1, 85748 Garching, Germany}
\author[0000-0001-9336-2825]{S. E. de Mink}
\affiliation{Max Planck Institute for Astrophysics, Karl-Schwarzschild-Strasse 1, 85748 Garching, Germany}
\affiliation{Anton Pannekoek Institute for Astronomy, University of Amsterdam, Science Park 904, 1098XH Amsterdam, The Netherlands}
\author[0000-0002-6718-9472]{M. Renzo}
\affiliation{Center for Computational Astrophysics, Flatiron Institute, New York, NY 10010, USA}
\author[0000-0001-7969-1569]{S. Justham}
\affiliation{School of Astronomy and Space Science, University of the Chinese Academy of Sciences, Beijing 100012, China}
\affiliation{Anton Pannekoek Institute for Astronomy, University of Amsterdam, Science Park 904, 1098XH Amsterdam, The Netherlands}
\affiliation{Max Planck Institute for Astrophysics, Karl-Schwarzschild-Strasse 1, 85748 Garching, Germany}
\author[0000-0002-7464-498X]{E. Zapartas}
\affiliation{IAASARS, National Observatory of Athens, Vas. Pavlou and I. Metaxa, Penteli, 15236, Greece}
\author[0000-0001-5228-6598]{K. Breivik}
\affiliation{Center for Computational Astrophysics, Flatiron Institute, New York, NY 10010, USA}
\author[0000-0001-9892-177X]{T. Callister}
\affiliation{Center for Computational Astrophysics, Flatiron Institute, New York, NY 10010, USA}
\author[0000-0003-1540-8562]{W. M. Farr}
\affiliation{Department of Physics and Astronomy, Stony Brook University, Stony Brook NY 11794, USA}
\affiliation{Center for Computational Astrophysics, Flatiron Institute, New York, NY 10010, USA}
\author[0000-0002-1590-8551]{C. Conroy}
\affiliation{Center for Astrophysics $|$ Harvard $\&$ Smithsonian,
60 Garden St., Cambridge, MA 02138, USA}

\begin{abstract}
Gravitational-wave (GW) detections are starting to reveal features in the mass distribution of double compact objects.
The lower end of the black hole (BH) mass distribution is especially interesting as few formation channels contribute here and because it is more robust against variations in the cosmic star formation than the high mass end.
In this work we explore the stable mass transfer channel for the formation of GW sources with a focus on the low-mass end of the mass distribution. We conduct an extensive exploration of the uncertain physical processes that impact this channel. 
We note that, for fiducial assumptions, this channel reproduces the peak at $\sim9\Msun$ in the GW-observed binary BH mass distribution remarkably well, and predicts a cutoff mass that coincides with the upper edge of the purported neutron star BH mass gap.
The peak and cutoff mass are a consequence of unique properties of this channel, namely (1) the requirement of stability during the mass transfer phases, and (2) the complex way in which the final compact object masses scale with the initial mass.
We provide an analytical expression for the cutoff in the primary component mass and show that this adequately matches our numerical results.
Our results imply that selection effects resulting from the formation channel alone can provide an explanation for the purported neutron star --BH mass gap in GW detections. This provides an alternative to the commonly adopted view that the gap emerges during BH formation.
\end{abstract}

\section{Introduction} \label{sec: intro}
Gravitational-wave (GW) events are revealing substructure in the mass
distribution of merging double compact objects
(\citealt{GWTC2_popPaper2021}, \citealt{GWTC3_popPaper2021}, cf. \citealt{Fishbach+2020_NSBH},  \citealt{TiwariFairhurst2021} and \citealt{Tiwari2022}). 
Understanding the origin of these features provides insight into the
physics of binary formation and evolution
\citep[e.g.,][]{Stevenson+2015,Fishbach+2017,Barrett+2018,Wysocki+2019,Fishbach+2020_massiveBBH,Vitale+2020,Doctor+2020,Belczynski+2020,Romero-Shaw+2021,Wong+2021}.
A better understanding of features in the mass distribution may enable
us to break the degeneracy between the observed source mass and
redshift from GW sources, which would provide a powerful cosmological
probe \citep[also known as `dark sirens' or `spectral sirens', 
e.g.,][]{Schutz1986,Farr+2019,Farmer+2019,MariaEzquiaga2022}. 
Additionally, redshift evolution of different parts of the mass
distribution can provide constraints on the cosmic star formation rate
from a completely new perspective
\citep[e.g.,][]{Vitale+2019,vanson+2021,chruslinska2022_review}.

At present it is difficult to take full advantage of the information that is contained within the mass distribution due to the uncertain origin of the compact object mergers.
Many channels have been proposed to explain the formation of double
compact objects (see the reviews from \citealt{Mapelli2020_review} and
\citealt{MandelFarmer2022}, and references therein). The mixing fraction between these formation channels is unclear \citep[e.g.,][]{Zevin2021,Wong+2021}. 
Moreover, large uncertainties in the evolution of massive stellar binaries lead to significant uncertainties in the predictions for the formation of GW sources, this is especially true for predictions from binary population synthesis models \citep[e.g.,][]{Abadie+2010,Dominik+2015,deMinkBelczynski2015,GiacobboMapelli2018,Tang+2020,Broekgaarden+2021a,Bavera+2021,Belczynski+2022}
It is therefore crucial to find predicted features in the source property distributions that are characteristic and unique to a single formation channel.\\

The lower end of the BH mass distribution (component masses of $\leq 15\Msun$) is the most promising site to reveal the origin of double compact objects for two reasons.
First, the low mass end of the binary black hole (BBH) mass distribution is least affected by the uncertainties in the metallicity-dependent star formation rate { \citep[][]{vanson+2022}}.
Second, only a few formation channels are relevant at the low-mass regime.
Only isolated binary evolution channels have been suggested to produce a global peak of the BH mass distribution at the low mass end \citep[][]{Belczynski+2016Natur,GiacobboMapelli2018,Giacobbo+2018,Wiktorowicz+2019,Belczynski+2020,Tanikawa+2022}.
The mass distributions from other channels, such as hierarchical formation \citep{Askar2017,Rodriguez:2019huv,Antonini+2019,Fragione+2020,FragioneSilk2020,AntoniniGieles2020}, chemically-homogeneous evolution \citep[CHE; e.g.\ ][]{de-Mink+2009,MandelMink2016,Marchant+2016,Riley2021}, population III binaries \citep[e.g.\ ][]{Marigo:2001,Belczynski:2004popIII,Kinugawa:2014,Inayoshi2017} and binaries merging in the disks of active galactic nuclei \citep[e.g.\ ][]{Baruteau2011,Bellovary2016,Leigh2018,Yang+2019,Secunda2019,McKernan2020} are expected to peak at masses above $20\Msun$.  
\cite{Antonini+2022} furthermore show that the globular cluster channel under-predicts the observed rate of BBH mergers at the low mas end (around $10\Msun$) by about two orders of magnitude.
Less confusion about the dominant formation channel also makes the low mass end one of the most promising sites to distinguish any astrophysical redshift evolution of the mass distribution from cosmological evolution  \citep[e.g., ][]{MariaEzquiaga2022}. \\

The latest catalogue of GW events has revealed two new features at the low end of the mass distribution of merging binary black holes (BBH). We expect that these findings are most likely two sides of the same coin and hence need to be jointly investigated.
First, the distribution of more massive components of merging BBH systems peaks at approximately $9\Msun$ \citep{GWTC3_popPaper2021,Li+2021,Veske+2021,Tiwari2022,Edelman+2022}. From hereon, we will use `primary' (secondary) to describe the more (less) massive component of double compact objects.
This feature at $9\Msun$ forms the \textit{global} peak in the primary
BH mass distribution \citep[][]{Tanikawa+2022}, which implies that the merger rate of $3\Msun$ BHs is lower than the rate of 9\Msun BHs.
This is surprising, because lower mass BHs are expected to form from lower-mass progenitor stars \citep[cf.][]{WoosleyHegerWeaver2002,Spera+2015,Woosley+2020}, which are heavily favoured by the initial mass function \citep[e.g.,][]{Kroupa2001}. 
Second, there is tentative evidence for a relative dearth of merging BBH observations with component masses between 3\Msun and 5\Msun.
Although at the time of writing, definitive statements about this dearth are hindered by the scarcity of detections in this mass range, \cite{Farah+2021}, \cite{YeFishbach2022} and \cite{Biscoveanu+2022} find that models for the mass distribution as observed in GW \textit{with} a gap are preferred over models \textit{without} a gap. 
If such a gap is allowed in the model, \cite{Farah+2021} find that a `rise' from this gap is expected between about 4.5 and 8.5\Msun (see the blue band in Figure \ref{fig: fiducial mass dist}). Future detectors will decisively probe the existence and location of a low-mass gap in the observations \citep[e.g.,][]{Baibhav+2019}.

Several works have suggested a gap in the remnant mass distribution between the most massive neutron stars (NSs) and the least massive BHs as an explanation of the dearth of low-mass BHs observed in gravitational waves  \citep[e.g.,][]{Zevin+2020,Farah+2021,Olejak+2022}. 
This notion of a `NS--BH mass gap' was originally inspired by observations of X-ray binaries, and has been a topic of active debate for over a decade \citep[e.g.,][]{Bailyn+1998,Ozel+2010,Farr+2011,Kreidberg+2012,Casares+2017,WyrzykowskiMandel2020}.
The discussion ranges from the observational selection biases that could create the \textit{appearance} of a mass gap \citep[e.g.,][]{Jonker+2021,Siegel+2022,Liotine+2022}, to the theoretical explanation under the assumption that the mass-gap is real (e.g., a fallback mechanism as proposed by \citealt{Fryer+2012}, \citealt{Fryer+2022}, or failed supernova as proposed by \citealt{Kochanek2014},\citealt{Kochanek2015}). 

Alternatively, it could be that there is an \textit{evolutionary} selection bias at play that excludes the formation of merging double compact objects with component masses of about 3-5\Msun.
In this case, features in the mass distribution could be a telltale sign of the dominant formation channel.

The channels that are expected to dominate BBH formation with low component masses are the stable mass transfer channel \citep[e.g.,][]{van-den-Heuvel+2017,Inayoshi2017,Bavera+2021,Marchant2021,GallegosGarcia2021,vanson+2021}, and the `classical' common-envelope channel \citep[or CE channel, e.g.\  ][]{Belczynski+2007,PostnovYungelson2014,Belczynski+2016Natur,vignaGomez+2018}. These channels are both forms of isolated binary evolution, and are distinguished based on whether the binary experiences common envelope evolution (CE channel) or only stable mass transfer (stable channel in short from now on).
Recent work suggests that the contribution of the CE channel to the BBH merger rate might be overestimated in rapid population synthesis simulations \citep[e.g.,][]{Pavlovskii2017,Klencki2021,Marchant2021,GallegosGarcia2021,Olejak+2021}. They argue that many of the systems that are assumed to lead to successful CE ejection in rapid population synthesis codes should instead either lead to stable mass transfer or a stellar merger.
This has caused the stable mass transfer channel to receive renewed attention as a plausible dominant channel for the formation of merging BBHs \citep[e.g.,][]{Shao_Li2022,Briel+2022}.

\subsection{Motivation for this work}
\begin{figure}
\centering
\includegraphics[width=0.49\textwidth]{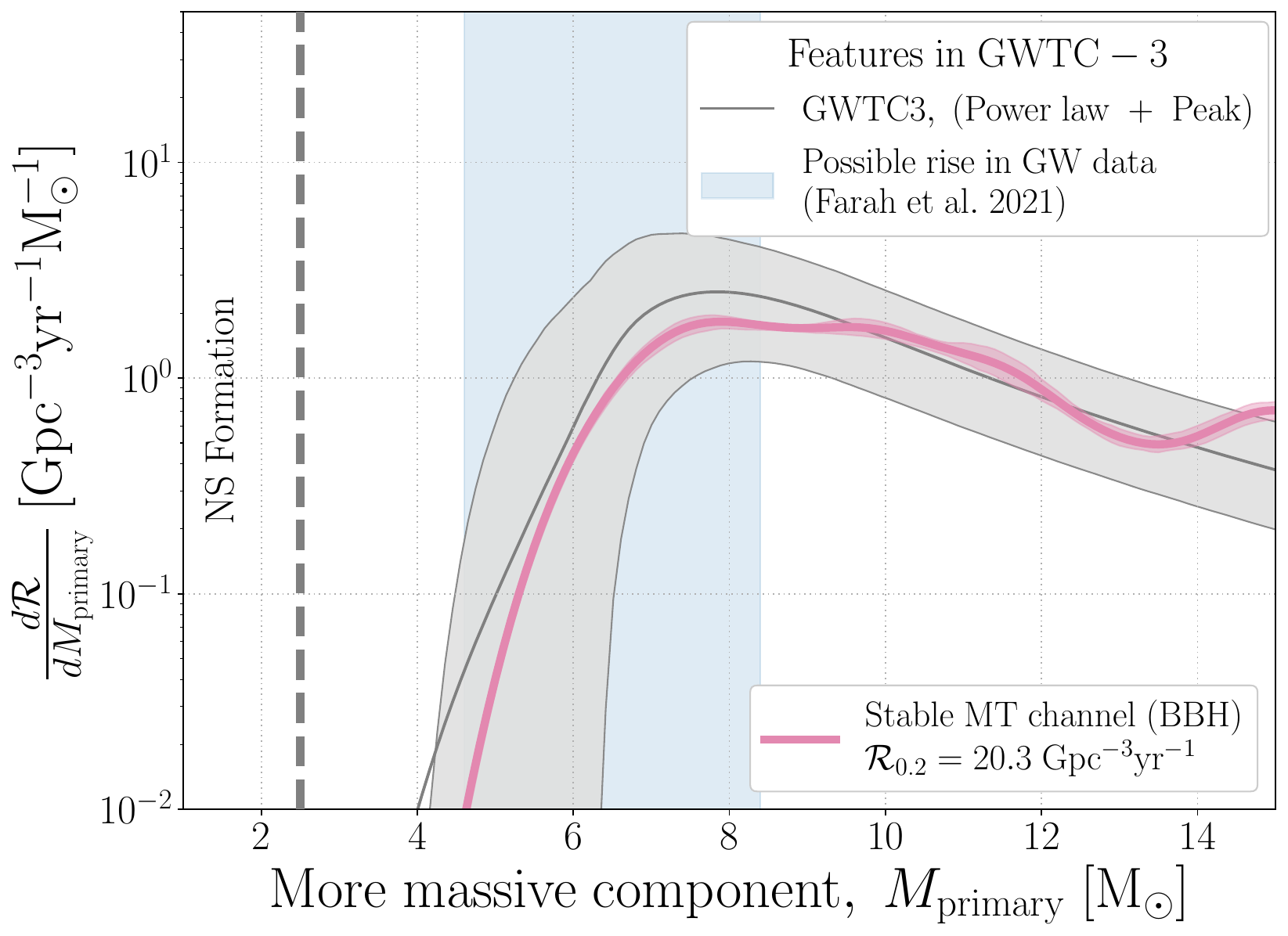}
\caption{Intrinsic distribution of primary masses from BBH merging at redshift 0.2. We show the fiducial predictions for the stable channel in the pink kernel density distributions. The light-shaded area shows the $90\%$ sampling uncertainty as obtained from bootstrapping.
The total merger rate of BBHs at $z=0.2$ is annotated in the legend. The power law + peak model from \cite{GWTC3_popPaper2021} is shown in grey, light grey bands show the 90\% credible intervals. We indicate a tentative rise observed in the GW data with filled blue (see text).
We see that the local rate and the location of the peak at the low mass end of the primary BH mass distribution can be explained remarkably well by the stable channel under our fiducial assumptions.
\label{fig: fiducial mass dist} }
\end{figure}

The inspiration for the work in this paper is shown in Figure \ref{fig: fiducial mass dist}.
This Figure was produced shortly after the release of the third GW catalogue  \citep[GWTC-3][]{GWTC3,GWTC3_popPaper2021}, using \COMPAS version \texttt{v02.26.03} with the exact same settings as the fiducial model for isolated binary formation from \cite{vanson+2021}, i.e., this is not optimised to match the observations.
In pink we show the fiducial predictions from the stable channel. The characteristic of this channel is that every mass transfer episode throughout the binary evolution is dynamically stable, and no common envelope occurs.
The main reason for the orbit to shrink in this channel is loss of mass with high specific angular momentum from the vicinity of the lower mass companion.

There is a striking similarity between the GW inferred BBH mass distribution and our predictions for the stable channel, shown in Figure \ref{fig: fiducial mass dist}.
This model reproduces both a) the dearth of merging primary BH masses between 2.5-6\Msun, and b) a peak around 8-10\Msun.
It also matches the local intrinsic rate of BBH mergers.
{As shown in \citet{vanson+2022}, the \textit{location} of features could in particular serve as sign posts of the underlying physics.}
However, at present, it is not clear whether this resemblance is coincidental given the uncertainties that plague population synthesis modeling \citep[see e.g.,][]{Dominik+2015,deMinkBelczynski2015,GiacobboMapelli2018,Broekgaarden+2021a,Belczynski+2022} and the significant model dependence involved in the GW-inference of the mass distribution \citep[e.g.,][]{GWTC3_popPaper2021}. That is, it could be that we are getting the right result for the wrong reasons.

To better understand \textit{why} this model provides a good fit, we investigate the stable mass transfer channel in more detail in this work.
In particular, we set out to explore 1) why the stable channel experiences a sharp rise that turns into a peak around 6\Msun, 2) the physical processes that dominate the shape of the mass distribution, and 3) how robust this feature is against variations.
We find that the stable channel leads to a cutoff in the primary mass, \Mmm for BBH and BHNS systems.
Adopting a set of simplifying assumptions, we analytically express this minimum mass as a function of birth mass ratio, and determine the main uncertainties in the physical assumptions that dictate the minimum value of \Mmm. {We discuss how this cutoff mass affects the location of the peak of the BBH mass distribution, while it could also} lead to a decrease or even a gap in the mass distribution of \Mmm that follows from GW events, without the need for a gap in the supernova remnant mass function.  \\

The remainder of this paper is structured as follows: we define the key parameters and assumptions needed to describe typical evolution through the stable channel, and show how these lead to a cutoff mass in Section \ref{sec: definitions}.
In Section \ref{sec: validation and behaviour} we compare our analytically derived minimum to numerical simulations and confirm that the physics variations considered lead to a comprehensive understanding of the minimum mass. We furthermore compute the corresponding mass distribution for every variation considered.
We explore the effect of a more complex supernova remnant mass function and of mass loss into a circumbinary disc in Section \ref{ss: dMsn and gamma}. 
Finally, we discuss implications of constraints on the primary mass as expected for the `stable mass transfer channel' in Section \ref{sec: discussion}, and we summarise our findings in Section \ref{sec: conclusions}.

\section{Analytic approximation of the stable mass transfer channel \label{sec: definitions} }
In Section \ref{ss: evolution params} we describe the typical evolution of a binary through the stable channel in chronological order. We describe the key evolutionary steps in terms of uncertain physics parameters and explain our adopted analytical assumptions.
The parameters discussed throughout this section are shown in Figure \ref{fig: cartoon}, which depicts the key evolutionary steps of the stable channel.
In Section \ref{ss: derivation} we investigate constraints on the masses that follow from this channel.
%

\subsection{The evolutionary steps of the stable channel \label{ss: evolution params}}
\begin{figure}
\includegraphics[trim=10.5cm 0cm 5.5cm 0cm,clip=true,width=0.47\textwidth]{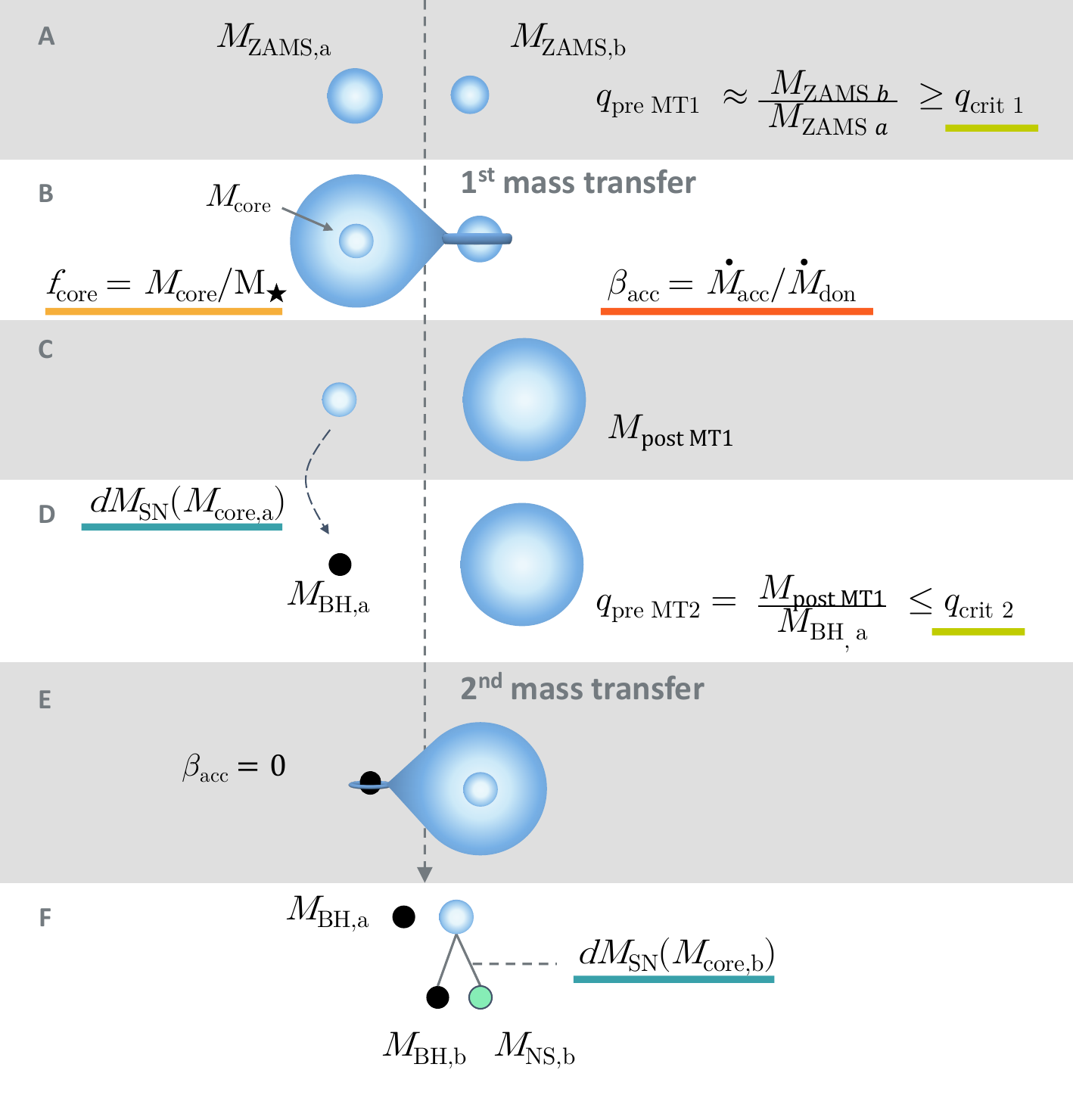} 
\caption{Cartoon depiction of the stable mass transfer channel, including the most relevant parameters. See Section \ref{ss: evolution params} and Table \ref{tab: fid values} for an explanation of the parameters. \label{fig: cartoon}}
\end{figure}

At the zero age main sequence (ZAMS, step A in Figure \ref{fig: cartoon} ) we define masses \MZAMSI and \MZAMSII for the respectively more and less massive binary component at the onset of H-burning. Throughout this work, we will refer to these components using the subscripts $a$ and $b$ accordingly.

The more massive star evolves on a shorter timescale and will typically overflow its Roche lobe first. We will refer to this as the first mass transfer event (step B in Figure \ref{fig: cartoon}). We assume that the donor star loses its complete envelope,
which implicitly assumes a well defined core-envelope structure, typical for post main sequence mass transfer. The relevant type of mass transfer is known as Case B mass transfer, which is the most common type of binary interaction especially for increasing metallicity \citep[e.g.,][]{vandenHeuvel1969,deMink+2008,Renzo+2019}. We will discuss the effects of this assumption in Section \ref{ss: disc. caveats}.

We define the core mass fraction, \fcore, as the fraction of the ZAMS mass that ends up in the He core mass ($\Mcore$) at the end of the main sequence,  i.e.,  $\fcore~=~\Mcore/M_{\rm{ZAMS}} $.
The fraction of mass lost by the donor star will be $1 - \fcore$.
We assume a fraction \betaacc of the transferred mass will be accreted by the companion star. We will refer to this as the mass transfer efficiency.
We assume that any mass lost from the system during a stable mass transfer event will carry away the specific angular momentum of the accretor \citep[also known as isotropic reemission; e.g.,][]{Soberman+1997}.

At step C in Figure \ref{fig: cartoon}, the initially more massive star has become a helium star, and the initially less massive star is still a main sequence star with new mass $\MtildeII=\MZAMSII+ \betaacc \MZAMSI(1-\fcore)$. In the latter approximation we neglect wind mass loss. We assume that the initially more massive star will collapse to form a compact object (step D). This is typically a BH for the systems we consider and we denote its mass as \MbhA.
Note that this will not necessarily be the more massive compact object.
Not all the mass of the core will end up in \MbhA. Part could be lost during the supernova (SN), and part will be lost to stellar winds. The SN mass loss is expected to become particularly important for low-mass BHs.
We parameterize this mass loss as
$\dMsn \equiv M_{\mathrm{core,1}} - \MbhA$ \citep[cf.\ ``top down''
approach in][]{Renzo+2022}. Because both winds and SN mass loss are expected to be mass dependent, \dMsn is also mass dependent.
Here, we assume mass lost from the system carries away the specific angular momentum of the donor (i.e.\ `Jeans mode').

A second mass transfer phase occurs as the initially less massive star evolves off the main sequence and overflows its Roche lobe (step E).
Typically accretion on BHs is limited by radiation pressure in the accretion flow, which leads to very low accretion rates (i.e., Eddington limited accretion). Hence we adopt $\betaacc = 0$ during the second mass transfer phase.
Finally, the initially less massive component collapses to form either a BH or a NS. We again parametrise the difference between the core mass and final remnant mass with \dMsn (step F).

In this work, unless stated otherwise we define mass ratios as the initially less massive over the initially more massive binary component. Hence $\qzams\equiv \MZAMSII/\MZAMSI$. The mass ratios right before the first and second mass mass transfer phases are thus respectively $\qpremtI = \MZAMSII/\MZAMSI$ and $\qpremtII = \MtildeII/\MbhA$.

To determine the dynamical stability of mass transfer, we approximate the response of the Roche radius to mass lost, $\zeta_{\mathrm{RL}} \equiv d\ln R_{\mathrm{RL}} / d\ln M_{\star} $, and compare this to an approximation of the adiabatic response of the donor star to mass loss, $\zeta_{\star} \equiv d\ln R_{\star} / d\ln _{\star}$ \citep[see e.g.,][]{Soberman+1997, COMPAS_method}.  Mass transfer is assumed to be stable as long as $\zeta_{\mathrm{RL}} \leq \zeta_{\star}$.
The value of $\zeta_{\star}$ is determined by the stellar structure of the donor in the adiabatic approximation \citep[e.g.,][]{Ge+2015,Ge+2020}. Throughout the rest of this work, we adopt $\zeta_{\star} = \zetaeff = 6.0$ as our reference value for Hertzsprung-gap donor stars (these are subject to the delayed dynamical instability, for which see
 \citealt{Hjellming+Webbink1987}).
$\zeta_{RL}$ is a function of \betaacc, and the mass ratio between the accretor and donor ($q= M_d/M_a$). 
The dependence of  \zetaeff on \betaacc, and the mass ratio between the accretor and the donor, is shown in Figure 4 of \cite{Soberman+1997}. For clarity, we also show this dependence for different values of \betaacc in Appendix \ref{app: zeta}.

The requirement of mass-transfer stability leads to a limit on the mass ratio between the accretor and donor. We will refer to these critical mass ratios as \qcritI and \qcritII for the first and second mass transfer phase, with $\betaacc~=~0.5$ and $\betaacc~=~0$ respectively.
The mass ratio right before the first mass transfer phase is $\qpremtI~=~M_{\rm b}/M_{\rm a}$, which we approximate with $\qpremtI~=~\MZAMSII/\MZAMSI$ in our analytical approximation. Since, at this point, the initially more massive star is overflowing its Roche-lobe, mass transfer will be dynamically stable as long as
$\MZAMSII/\MZAMSI = M_{\rm accretor}/M_{\rm donor} \geq \qcritI$.
Similarly, right before the second mass transfer, the mass ratio is defined as $\qpremtII =  M_{\rm b}/M_{\rm a} = M_{\rm{post MT1}}/\MbhA = M_{\rm donor}/M_{\rm accretor} \leq \qcritII$.

\subsection{Derivation of low-mass cutoff for primary components \label{ss: derivation} }
The main objective of this work is to understand constraints on the allowed compact-object masses at the low end of the mass distribution for the stable mass transfer channel.
The characteristic constraint of the stable mass transfer channel is that both the first and the second mass transfer phases must be stable. We start from the constraint on the second mass transfer phase, as we find that it is particularly decisive for the final masses involved.
This leads to an inequality between the mass ratio of the system at the onset of the second mass transfer phase and \qcritII,

\begin{equation}
\qpremtII = \frac{\MtildeII}{\MbhA} \leq \qcritII ,
\label{eq: q pre mt 2}
\end{equation}

\noindent where \qcritII is the critical mass ratio during the second mass transfer phase (i.e.\ assuming $\betaacc = 0$, see Section \ref{ss: evolution params}).
\MtildeII is the mass of the initially less massive star post mass accretion from the first mass transfer event.
We can approximate this as

\begin{equation}
\MtildeII =  \MZAMSII + \MZAMSI   \betaacc(1 - \fcore  ) ,
\label{eq: m tilde 2}
\end{equation}

\noindent and \MbhA as

\begin{equation}
\MbhA = \fcore   \MZAMSI   - \dMsn .
\label{eq: mbh1}
\end{equation}

\noindent Rewriting  Equation \ref{eq: q pre mt 2} using Equations \ref{eq: m tilde 2}, \ref{eq: mbh1} and $\qzams=\MZAMSII/\MZAMSI$ gives:

\begin{equation}
\frac{\qzams + \betaacc(1 - \fcore)}{\fcore  - \frac{\dMsn}{\MZAMSI} } \leq \qcritII .
\label{eq: rewrite}
\end{equation}

In this work, we are specifically interested in placing a lower bound on the possible masses of BBH and BHNS systems formed through the stable mass transfer channel.
At this point, the only explicit mass dependence left is \MZAMSI. However, both \fcore, and \dMsn implicitly depend on \MZAMSI. In order to find a lower bound on \MZAMSI,  we would like to make these dependencies explicit.

In general, \fcore is expected to increase with mass. It is however reasonable to adopt an approximately constant value for \fcore as long as the \MZAMSI range of interest is not too large. 
This is the case for the range of ZAMS masses relevant for producing the lowest mass BHs in our simulations. For $\MZAMSI \approx 20-40\Msun$, stellar evolution tracks in \COMPAS lead to core mass fractions of effectively $\fcore \approx 0.3 - 0.34$ \citep[which is a result of the assumptions in][on which the \COMPAS code was based]{Pols+1998}.
Hence from here on we continue using the simplification that is constant at $\fcore=0.34$ (though see appendix \ref{app: alt minimum mass} for an alternative scenario).

In reality, \dMsn is a complicated function that depends on both the structure of the core at the moment of core collapse, as well as on the dynamics of the collapse, bounce and shock propagation. However, in general we expect that lower mass cores more easily lead to a successful explosion, and hence lead to more mass loss, than higher mass cores \citep[e.g.,][]{Fryer+2012,Muller+2018}.
For our reference model we adopt the `Delayed' model from \cite{Fryer+2012}, which is a continuous function that maps CO core masses to final remnant masses. This allows us to express \dMsn as a linear function of the core mass;
\begin{equation}
    \dMsn(\Mcore) =
    \begin{cases}
        \asn \Mcore + \bsn &  \Mcore \leq M_{\mathrm{thresh} } \\
        0 &  \Mcore > M_{\mathrm{thresh} } . \\
    \end{cases}
    \label{eq: dMsn}
\end{equation}
Here $M_{\mathrm{thresh}}=14.8\Msun$ is the threshold core mass above which we assume full fallback occurs, and $\asn=-0.9$ and $\bsn=13.9$ are obtained through a linear fit to our reference model (see also in Table \ref{tab: fid values}).
For \dMsn we approximate the core mass as $\Mcore~=~\fcore\MZAMSI$, with \fcore constant.

Going back to Equation \ref{eq: rewrite}, we can now explicitly write all terms that depend on \MZAMSI on one side of the equation

\begin{equation}
    \frac{ \qcritII\fcore  - \betaacc(1 - \fcore) -\qzams  }{\qcritII} \geq \frac{\dMsn(\Mcore)}{\MZAMSI} , \\
\end{equation}

\noindent which we can re-write to

\begin{equation}
    \boxed{
    \MZAMSI \geq
     \frac{\bsn \qcritII}{ \qcritII \fcore (1 - \asn) -\betaacc(1 - \fcore) - \qzams }.
    }
\label{eq: minMzams dep Msn}
\end{equation}
So far, we have only used the mass transfer stability constraint from the second mass transfer phase. The requirement that the first mass transfer must be stable also places a constraint on the minimum allowed value for $\qzams \in[q_{\rm crit,1}, 1]$.
Hence, we can derive a cut-off mass for \MZAMSI by adopting $\qzams = \qcritI$.

Equation \ref{eq: minMzams dep Msn} implies that the minimum ZAMS mass that can lead to double compacts objects through the stable channel, is determined by the physics parameters that are relevant to mass transfer stability at the first and second mass transfer phase. These parameters include \qcritI and \qcritII, but also parameters determining the mass ratio at mass transfer, namely \betaacc, \fcore and $\dMsn(\asn,\bsn)$.

We can use  Equation \ref{eq: minMzams dep Msn} to further derive a minimum mass for each of the final compact objects.
For the remnant from the initially more massive star:
\begin{equation}
\min(\MbhA) = \fcore   \min(\MZAMSI) - \dMsn(\McoreA),
\label{eq: minMBH1}
\end{equation}

\noindent where $\dMsn(\McoreA)$ is a shorthand for  Equation \ref{eq: dMsn} at $\McoreA~=~\MZAMSI\fcore$.
Similarly, for the remnant from the initially less massive star;
\begin{equation}
    \min(\MbhB) = \fcore   \min(\MtildeII) - \dMsn(\McoreB) ,
\label{eq: minMBH2}
\end{equation}

where $\dMsn(\McoreB)$ is Equation \ref{eq: dMsn} at $\McoreB~=~\MtildeII\fcore$.

Using  Equation \ref{eq: q pre mt 2}, we can constrain $\min(\MtildeII)$ as
$$
\min(\MtildeII)=~\qcritII \min(\MbhA).
$$

Finally, to compare with GW observations, we are interested in the BH that will form the more massive (primary) component of the double compact objects, since we cannot infer from GW if the primary descends from the initially more or less massive star.
Therefore we consider

\begin{equation}
    \boxed{
    \min(\Mmm) = \max\left\{\min(\MbhA), \min(\MbhB) \right\} .
    }
\label{eq: min MBH moreM}
\end{equation}

Equation \ref{eq: min MBH moreM} sets a minimum to the primary mass that can originate from the stable channel. It is an analytical function that depends on the initial condition \qzams, and the uncertain physics parameters \qcritII, \betaacc, \fcore and \dMsn ($\asn, \bsn$). See Table \ref{tab: fid values} for the reference values of these parameters as used in this work.

\begin{deluxetable*}{l l c c} 
\tablecaption{Physics parameters and their reference values. \label{tab: fid values}}
\tablehead{\colhead{Variable} & \colhead{Description} & \colhead{Ref. value} & \colhead{ {Explored variations} } }
\startdata
\betaacc            & \parbox{75mm}{Mass transfer efficiency: fraction of donated mass accreted by the companion star}                           & 0.5                & [0.0, 0.25, 0.5, 0.75, 1.0]                    \\ 
\zetaeff            & \parbox{75mm}{Response of donor star to mass loss $\zetaeff~\equiv~d\ln R_{\star}/d\ln M _{\star}$}                      & 6.0                & [3.5, 4.5, 5.5, 6.0, 6.5]                    \\ 
(\qcritI, \qcritII) & \parbox{75mm}{Effective critical mass ratio for stable mass transfer, using $\betaacc=0.5$ and $0$ respectively  (first and second mass transfer phase) }  & (0.25, 4.32)   &  \parbox{55mm}{[(0.41,3.03), (0.35,3.55), (0.30,4.06), (0.28,4.32), (0.26,4.58)]}             \\  %
\fcore              & \parbox{75mm}{Core mass fraction. }                                                                                        & 0.34               & [0.28, 0.31, 0.34, 0.38, 0.41]                    \\
\asn, \bsn          & \parbox{75mm}{Fit parameters for supernova mass loss \dMsn  (eq.   \ref{eq: dMsn}) }                                       & $-0.9, 13.9\Msun$  & varied prescription to \protect\cite{Fryer+2022}              \\
$M_{\rm{thresh}}$       & \parbox{75mm}{Boundary mass for full fallback  (eq.~\ref{eq: dMsn})}                                                 & $14.8\Msun$        & varied prescription to \protect\cite{Fryer+2022} 
\enddata
\tablewidth{0.5\columnwidth}
\end{deluxetable*}

\section{Results: Effect of the minimum mass for the stable channel \label{sec: validation and behaviour}}
In this section we discuss a comparison of our analytical results presented in Section \ref{sec: definitions} with numerical simulations. For this we adopt a reference model that is very similar to the fiducial model in \cite{vanson+2021}, presented also in Figure \ref{fig: fiducial mass dist}. Below, we will shortly describe the differences. We refer the reader to methods section of \cite{vanson+2021} for a more detailed description of the remainder of adopted physics parameters.

\begin{figure*}
\centering

\includegraphics[trim=0.8cm 0.7cm 2cm 2cm,clip=true,width=0.46\textwidth]{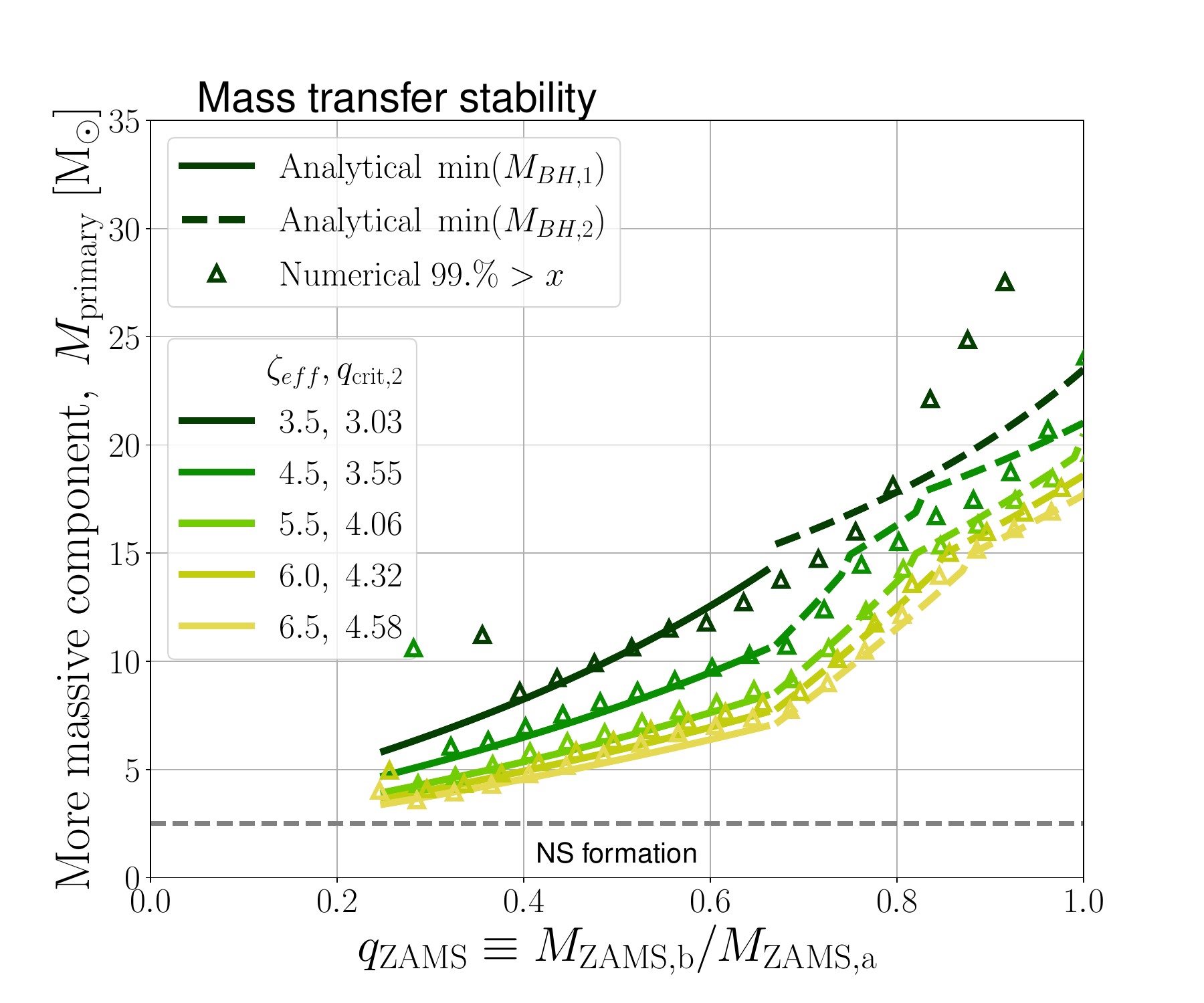}
\includegraphics[width=0.45\textwidth]{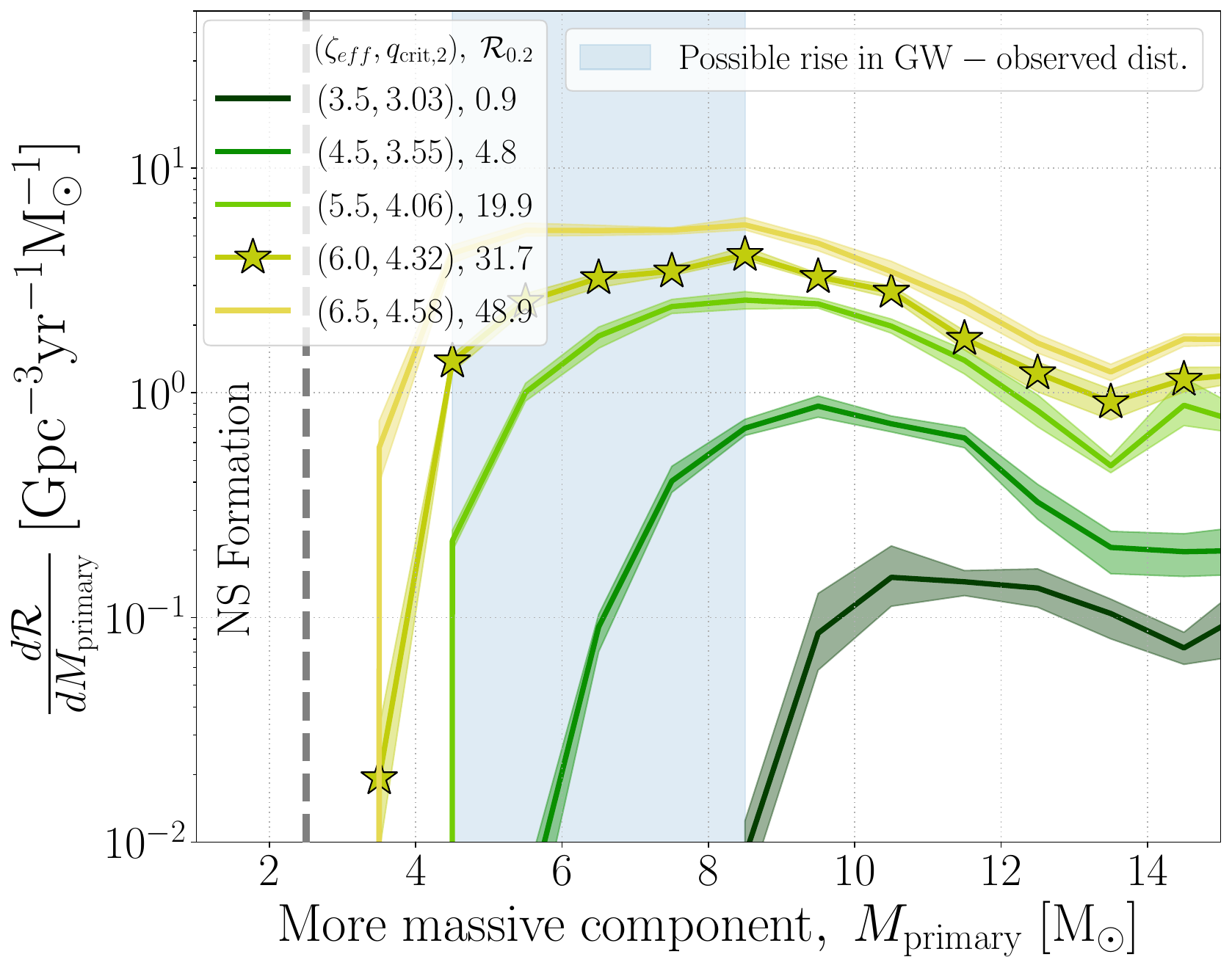}
\includegraphics[trim=0.8cm 0.7cm 2cm 2cm,clip=true,width=0.46\textwidth]{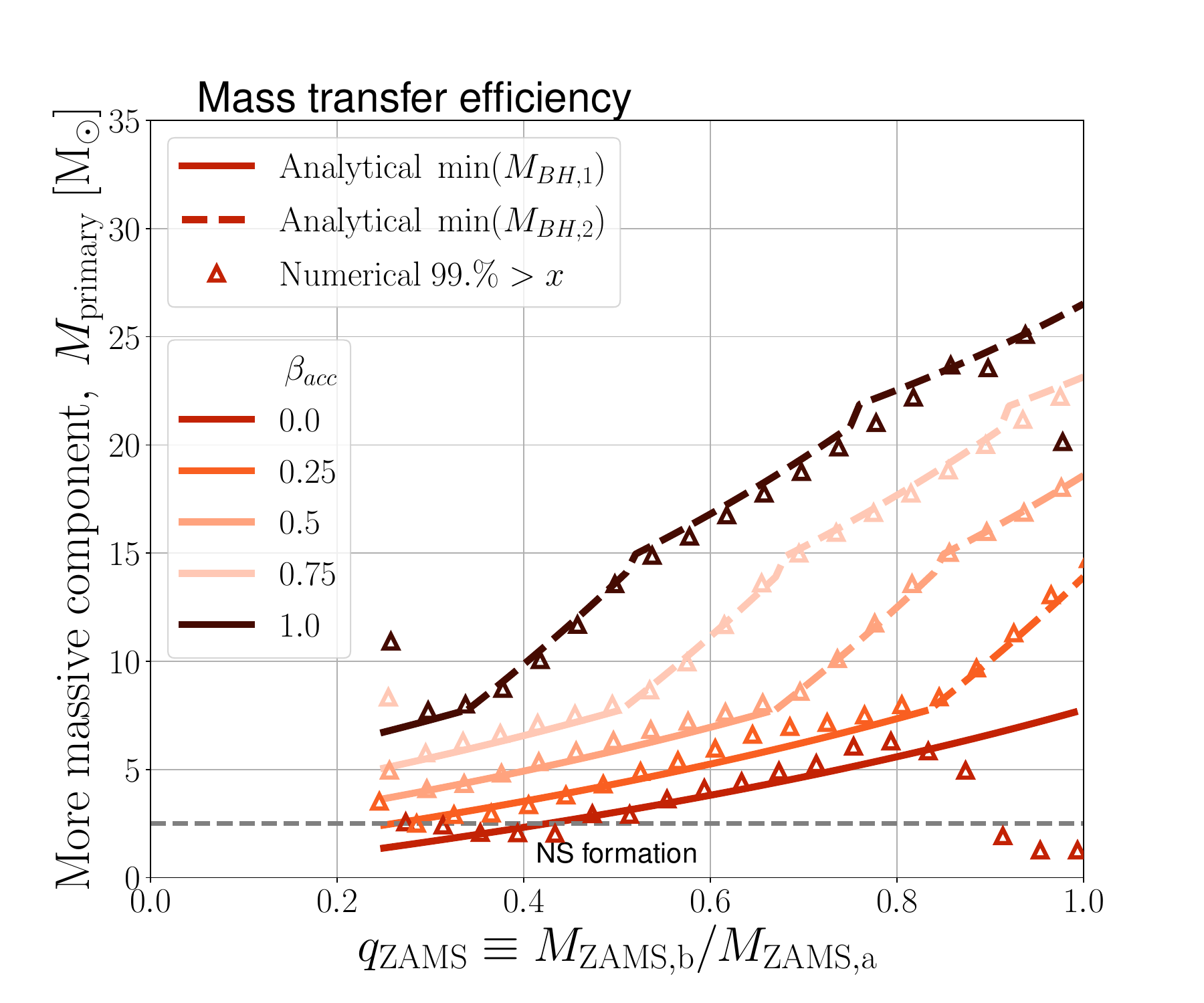}
\includegraphics[width=0.45\textwidth]{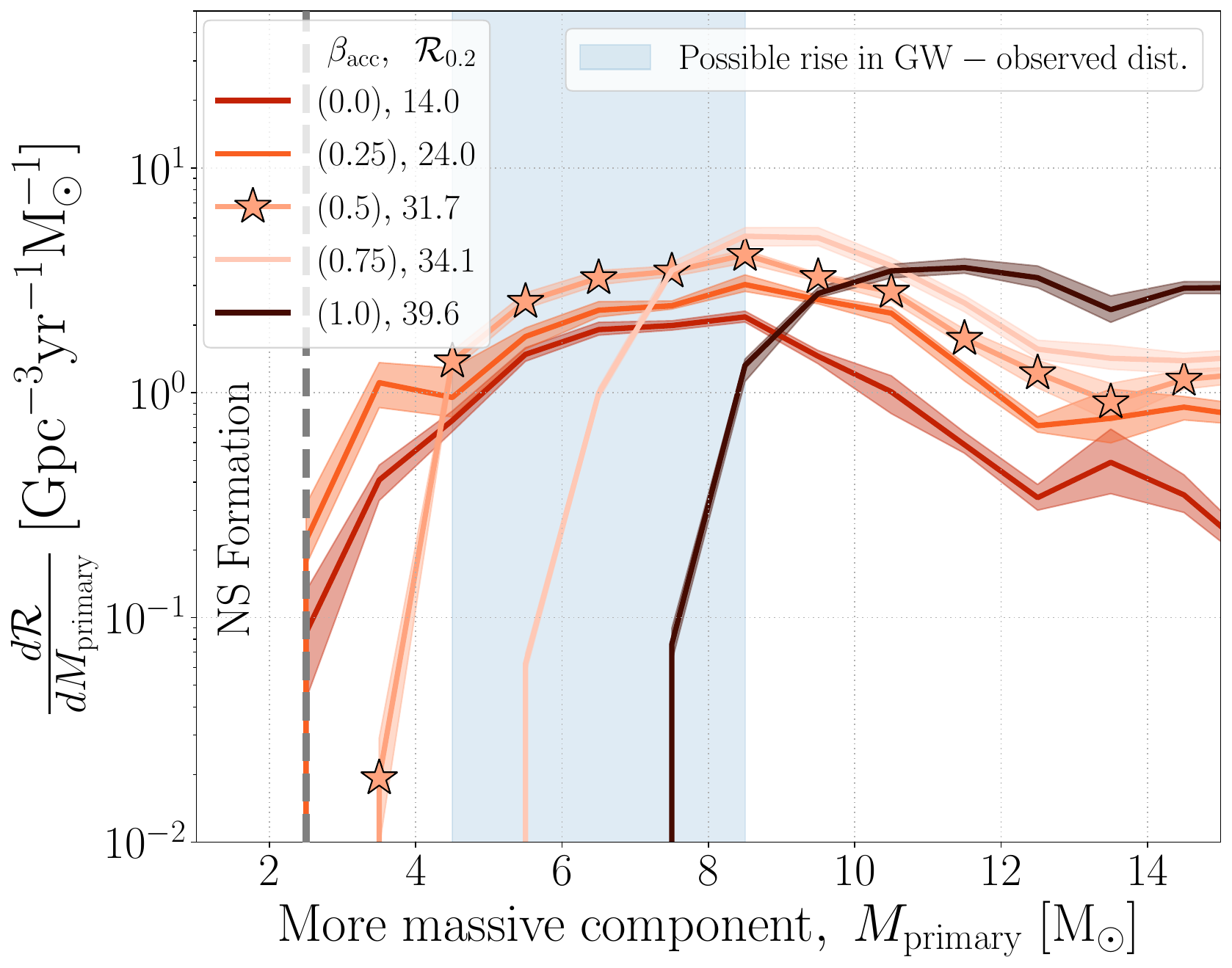} 
\includegraphics[trim=0.8cm 0.7cm 2cm 2cm,clip=true,width=0.46\textwidth]{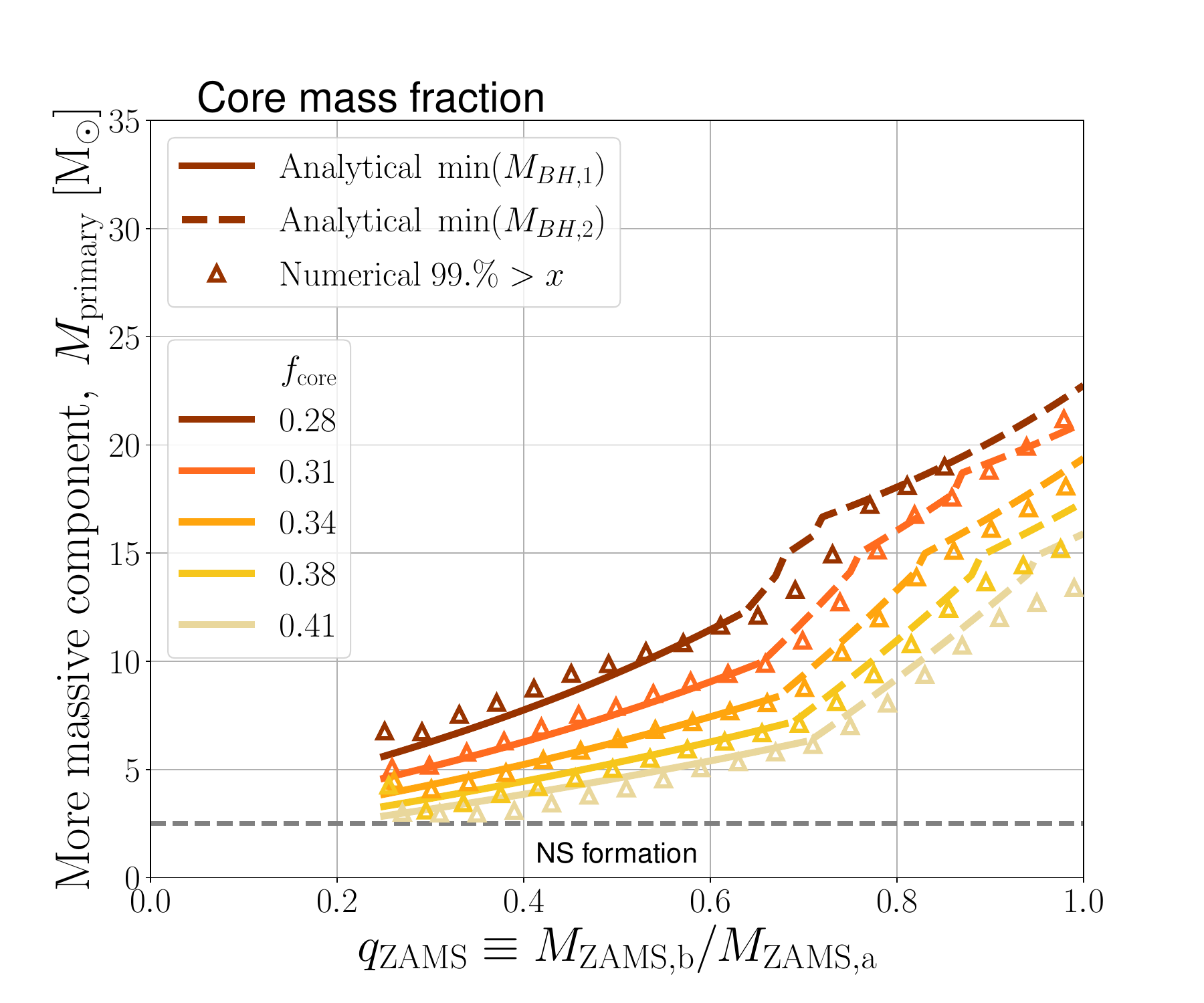} \includegraphics[width=0.45\textwidth]{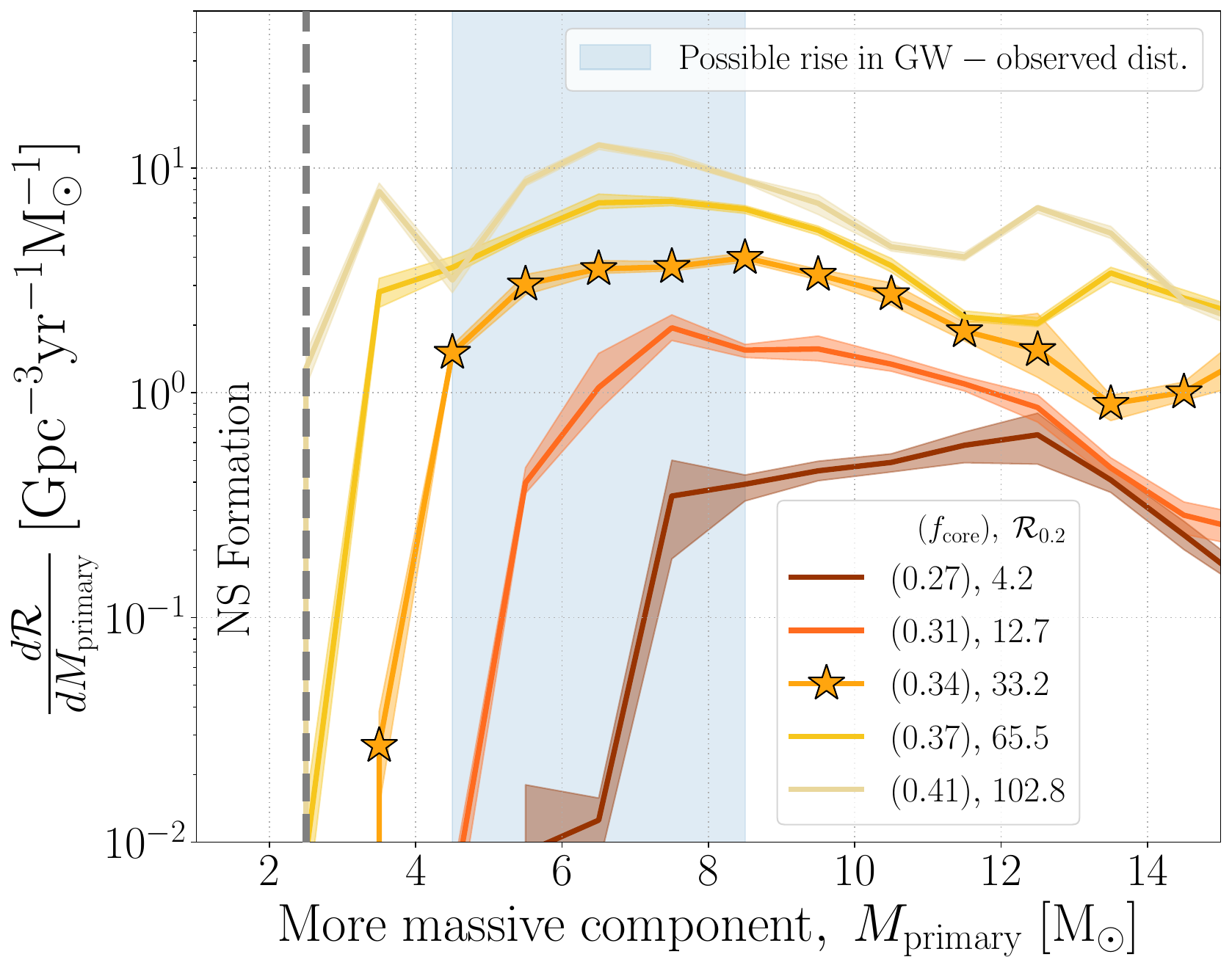}

\caption{Model predictions for the masses of BBH and BHNS systems formed through the stable channel.
\textbf{Left column:} \minmbh as a function of the ZAMS mass ratio \qzams. Lines show our analytical prediction from  Eq. \ref{eq: min MBH moreM}. Solid (dashed) lines indicate that \Mmm comes from \MbhA (\MbhB), described by  Eq. \ref{eq: minMBH1} (\ref{eq: minMBH2}).
Triangles show results from numerical simulations; $99\%$ of the simulation has a mass \Mmm larger than that value for bins in \qzams of 0.02.
\textbf{Right column:} Histogram of \Mmm for BBH and BHNS from the stable channel for bins in \Mmm of 1\Msun. The total rate at redshift 0.2 is annotated in the legend. Star markers indicate the reference model (Table \ref{tab: fid values}). Light-shaded areas show the $90\%$ sampling uncertainty, obtained by bootstrapping.
We show variations in the stability criteria (\zetaeff and \qcritII, top), the mass transfer efficiency (\betaacc, middle), and the core mass fraction (\fcore, bottom).
This shows that the analytically derived minimum can explain the numerical results well.
It furthermore displays how the cutoff mass in the stable channel leads to a dearth of BBH and BHNS systems with low primary masses for most variations.
\label{fig: validation and mass dists} }
\end{figure*}

Motivated by the variables in our analytical expression, we explore variations in the stability of the second mass transfer \qcritII, the mass transfer efficiency \betaacc and the core mass fraction \fcore. We discuss direct changes to the supernova remnant mass function in Section \ref{ss: dMsn and gamma}.
The varied physics parameters and their reference values are listed in Table \ref{tab: fid values}.
In contrast to the model in \cite{vanson+2021}, we adopt a fixed mass transfer efficiency value of $\betaacc=0.5$ as our reference value \citep{MeursVandenHeuvel1989,BelczynskiStarTrack+2008}, to enable a clear illustration of the effect described in Section \ref{sec: definitions}. The effect of adopting a mass transfer efficiency that varies with accretor properties is discussed in Section \ref{ss: disc. caveats}.
We adopt $\zetaeff = 6.0$ as our reference for radiative envelope donors with a clear core-envelope structure, compared to $\zetaeff = 6.5$ in \cite{vanson+2021}.
For $\zetaeff = 6.0$, the maximum mass ratio that leads to stable mass transfer $\qcritII \approx 4.32$ for fully non-conservative mass transfer, compared to 4.6 for $\zetaeff = 6.5$. Both values of \zetaeff are in agreement with the work of \cite{Ge+2015}.
The value of \fcore =0.34 is chosen as the best fit to our reference simulation.
Similarly, the values for \asn and \bsn are obtained from a fit to the difference between the pre-SN core mass and remnant mass as a function of the pre-SN core mass for our reference simulation.

In total we ran 25 variations on our reference model. Each simulation set contains  $10^7$ binaries run with version v.02.26.03 of the \COMPAS suite \citep{COMPAS_method}. To reduce sampling noise, we have sampled binaries using adaptive importance sampling \citep{Broekgaarden2019} optimising for BBH and BHNS mergers.

\subsection{Comparison to numerical data \label{ss: validation}}
For this analysis, we include all BBH and BHNS that have experienced exclusively stable mass transfer (i.e. we do not include chemically homogeneously evolving systems).
{We choose to show both BBH and BHNS, because our analytical prescription in Equation \ref{eq: min MBH moreM} does not require the outcome to be either a BBH or BHNS.}
We furthermore exclude binaries that never interact, or experience only one phase of mass transfer, since such systems are not expected to obey to our derived $\min(\Mmm)$, and because such systems are much too wide to form GW events.

We compare  Equation \ref{eq: min MBH moreM} to our grid of numerical simulations in the left column of Figure \ref{fig: validation and mass dists}. 
Triangles show where $99\%$ of each simulation has a mass \Mmm larger than that value, for bins in \qzams of width 0.02. We do not include bins with less than 10 samples.
Lines show our analytical prediction from  Equation \ref{eq: min MBH moreM}. Solid (dashed) lines indicate that \Mmm comes from \MbhA (\MbhB) and is described by  Equation \ref{eq: minMBH1} (\ref{eq: minMBH2}). Figure \ref{fig: validation and mass dists} shows that our analytical prediction of $\min(\Mmm)$, described by  Equation \ref{eq: min MBH moreM}, is in good agreement with the numerical data at almost every \qzams for all physics variation explored here.

The strongest deviations occur at two points.
For $\zetaeff =3.53$ (dark green line top right panel) we see that our prescription under-predicts the minimum primary mass from numerical simulations. This is effectively sampling noise: at low \zetaeff, we heavily reduce the window for stable mass transfer. Hence, for this variation, we barely sample any systems with high \qzams that do not experience unstable mass transfer.
Furthermore, at $\betaacc = 0$ and $\qzams \approx 1$ (bright red line middle right panel), we over-predict the minimum primary mass. We find that this is caused by nearly equal life-time of the two stars: in these cases, the initially more massive star has not yet finished the He-core burning phase when the initially less massive star evolves off the main sequence and overflows its Roche Lobe. This means that $\qpremtII = \MZAMSII/(\fcore \MZAMSI)$, which will be smaller than the assumed $\qpremtII = \MZAMSII/\MbhA$ in our analytical formula. Hence, the second mass transfer phase is more stable than our analytical formula predicts, and lower primary masses can be formed.

We note that for all variations, \minmbh increases with $\qzams$.
If there is a relation between \qzams and the final double compact object mass ratio \qfinal, then this implies a relation between \minmbh and the observed \qfinal.

{The absolute minimum \Mmm formed through the stable channel is found at \qzams=\qcritI. 
In other words, the stable channel will only contribute significantly to systems with $\qzams\geq\qcritI$.
Because \qcritI is a function of both \betaacc and \zetaeff (see Appendix \ref{app: zeta}), we expect that the minimum $\qzams$ at which the stable channel contributes significantly will also depend on \betaacc and \zetaeff. We see this effect in the top-left and middle-left panels of Figure \ref{fig: validation and mass dists}.
For lower \zetaeff, the minimum \qzams shifts to higher values because \qcritI increases.
That is, we only find systems with $\qzams\geq 0.25$ for \zetaeff=6.5, while for \zetaeff=3.5, this shifts to  $\qzams\geq 0.4$. 
Similarly, for \betaacc=0.0 systems with $\qzams\gtrsim 0.25$ contribute to the distribution while for \betaacc=1.0, $\qzams\gtrsim 0.33$.}

\subsection{Effect of minimum mass on mass distributions \label{ss: mass dist behaviour} }
{We show the distribution of \Mmm for merging BBH and BHNS in the right column of Figure \ref{fig: validation and mass dists}. Note that this is different from Figure \ref{fig: fiducial mass dist}, where we show only merging BBH. 
The reason for showing both BBHs and BHNSs is twofold. Firstly, we would like to confirm if the stable channel could lead to a dearth of low mass BHs that could be interpreted as a NS-BH mass gap (see Section \ref{sec: intro}). Excluding BHNS systems could unintentionally create an artificial dearth of low mass BHs. Second, we aim to explore and explain the behaviour of the stable channel. Hence, in order to investigate the effect of the minimum \Mmm on the resulting mass distribution, we integrate each of the physics variations as shown in the left hand panels of Figure \ref{fig: validation and mass dists}, over the metallicity-dependent star formation rate density as described in \cite{vanson+2021} and \cite{vanson+2022} (which is based on the approach of earlier work, e.g., \citealt{Dominik2013,Dominik+2015,Belczynski2016,Neijssel+2019,Broekgaarden+2021b}). 
To emphasise the steep features in the mass distribution, we use a histogram instead of a kernel density distribution to display the distribution of primary masses.}

We see that a higher cutoff mass can move the minimum primary mass to values that are significantly higher than the maximum NS mass.
{This affects the location of the peak of the mass distribution, while also  potentially opening up a gap between the most massive NS and the least massive BH}. Whether such a gap occurs is determined by the adopted physics variations. For many of our physics variations, the stable mass transfer channel is unable to form BBH or BHNS mergers with primary masses $\Mmm\sim 3$--$4\Msun$. Below we consider the effect of each physics variation on the primary mass in more detail. For completeness, we also show the chirp mass and final mass ratio distributions in Appendix \ref{app: other mass dists}. Throughout this section, we will refer to the combined rate of BBH and BHNS as \Rlocal. We include an overview of the individual BBH and BHNS rates as predicted by the stable channel in Appendix \ref{app: rates}.

\paragraph{Variations in the mass transfer stability \zetaeff}
Lower values of \zetaeff, and equivalently lower values of \qcritII, leave less room for stable mass transfer and severely restrict the window for stable mass transfer. Lower values of \zetaeff (darker green), lead to higher cutoff masses in \Mmm. A higher cutoff mass also shifts the peak of the mass distribution towards higher masses.
Less room for stable mass transfer furthermore significantly reduces the total merger rate for the stable channel (from $\Rlocal~\approx~49\Gpc^{-3}\yr^{-1}$ for $\zetaeff=6.5$ to $\Rlocal~\approx~0.9\Gpc^{-3}\yr^{-1}$ for $\zetaeff=3.5$ ). For $\zetaeff=6.5$ the stable mass transfer channel can form almost all primary BH masses, though primary black hole masses of about $3\Msun$ are still much less common than $\Mmm \sim 8\Msun$. For $\zetaeff=3.5$, the stable mass transfer only produces BHNSs and BBHs with primary masses above about $9\Msun$.
{We further note how \Mmm derives from the initially more massive component (\MbhA), for systems with $\qzams \lesssim 0.65$, while it derives from the initially less massive component (\MbhB) for $\qzams\gtrsim 0.65$, for every variation of \zetaeff (as can be seen in the upper left panel of Figure \ref{fig: validation and mass dists}). }

\paragraph{Variations in the accreted mass \betaacc}
Higher values of \betaacc significantly raise the minimum value of \Mmm at constant \qzams. Moreover, the slope of $\min(\Mmm)$ with \qzams increases for higher \betaacc. We understand this through the change in $M_{\rm{post MT1}}$. For larger \betaacc, $M_{\rm{post MT1}}$ will be larger, leading to larger \qpremtII, which leaves less room for stable mass transfer. This effect is more severe for $\qzams \sim 1$, since this implies a more massive companion star at ZAMS.
\betaacc influences whether \Mmm derives from \MbhA versus \MbhB. For $\betaacc = 1.0$, \Mmm almost always derives from the initially less massive star (except for $\qzams<0.35$, light pink line). For $\betaacc=0.0$, \Mmm is always \MbhA \citep[cf.][]{Broekgaarden+2022,ZevinBavera2022}.
For $\betaacc=1$, the distribution in \Mmm drops off steeply below about 8\Msun, while for $\betaacc=0$ there no real gap left in the mass distribution.
We again note how the location of the peak of the mass distribution is determined by the cutoff mass in \Mmm.

\paragraph{Variations in the core mass fraction \fcore}
The general behaviour of the core-mass fraction is similar to the effect of variations in the mass transfer stability: the peak of the primary mass distribution shifts to higher masses while the overall rate decreases.
Increasing the core mass fraction makes the second mass transfer phase more stable for constant values of \qzams.
This is because for higher \fcore, \qpremtII is lower and thus less likely to exceed \qcritII. $\qpremtII~=~M_{\rm{post, MT1}}/\MbhA$ is lower for higher \fcore both because \MbhA is more massive due to the higher core mass of the initially more massive star, and because $M_{\rm{post MT1}}$ is reduced since there is less envelope left to be accreted during the first mass transfer phase.
Hence, higher core mass fractions allow lower \Mmm to contribute to the stable mass transfer channel. Increasing the core mass fraction by 20\% ($\fcore\sim0.41$) with respect to our fiducial simulation causes the stable mass transfer channel to produce \Mmm with masses down to the NS limit of $2.5\Msun$. Moreover, this increases the rate to about $103 \Gpc^{-3} \yr^{-1}$.
Conversely, lowering the core mass fraction by 20\% to $\fcore\sim0.27$ lowers the rate to about $4 \Gpc^{-3} \yr^{-1}$, while only allowing $\Mmm\geq 7\Msun$.

\subsection{Variations in the SN mass loss and angular momentum loss \label{ss: dMsn and gamma}}
In this section we explore two further variations that are not captured by our simplified analytical model, while they are expected to significantly impact the mass distribution of merging double compact objects resulting from the stable channel.

\subsubsection*{Supernova remnant mass function \label{ss: SN variations}}
In Section \ref{ss: mass dist behaviour} we explore variations on all variables that appear in our analytical expression Equation \ref{eq: min MBH moreM}, except for the supernova mass loss \dMsn (Equation \ref{eq: dMsn}). The supernova mass loss is special, because variations in this function can cause a gap between BH and NS masses even in single stars, regardless of whether a double compact object forms (see also Section \ref{sec: intro}).

Here, we explore variations in the supernova remnant mass function by applying the new prescription from \cite{Fryer+2022}. In this prescription the remnant mass is a function of the carbon oxygen core mass at core-collapse,  $M_{\rm{crit}} = 5.75\Msun$; the lower boundary on the carbon oxygen core mass for BH formation (lower mass cores will form a NS) and \fmix, which describes the mixing growth time; higher \fmix corresponds to a more rapid growth of the convection.
Similar to \cite{Fryer+2022} and \cite{Olejak+2022} we explore variations between $\fmix=0.5$, which is closest to the `DELAYED' model in \cite{Fryer+2012}, and $\fmix=4.0$ which is most similar to the `RAPID' model in \cite{Fryer+2012}.
We apply BH kicks according to the `fallback' model from \cite{Fryer+2012}, where we adopt the proto-NS masses ($M_{\rm{proto}}$) from the DELAYED model.

\begin{figure}
\centering
\includegraphics[width=0.45\textwidth]{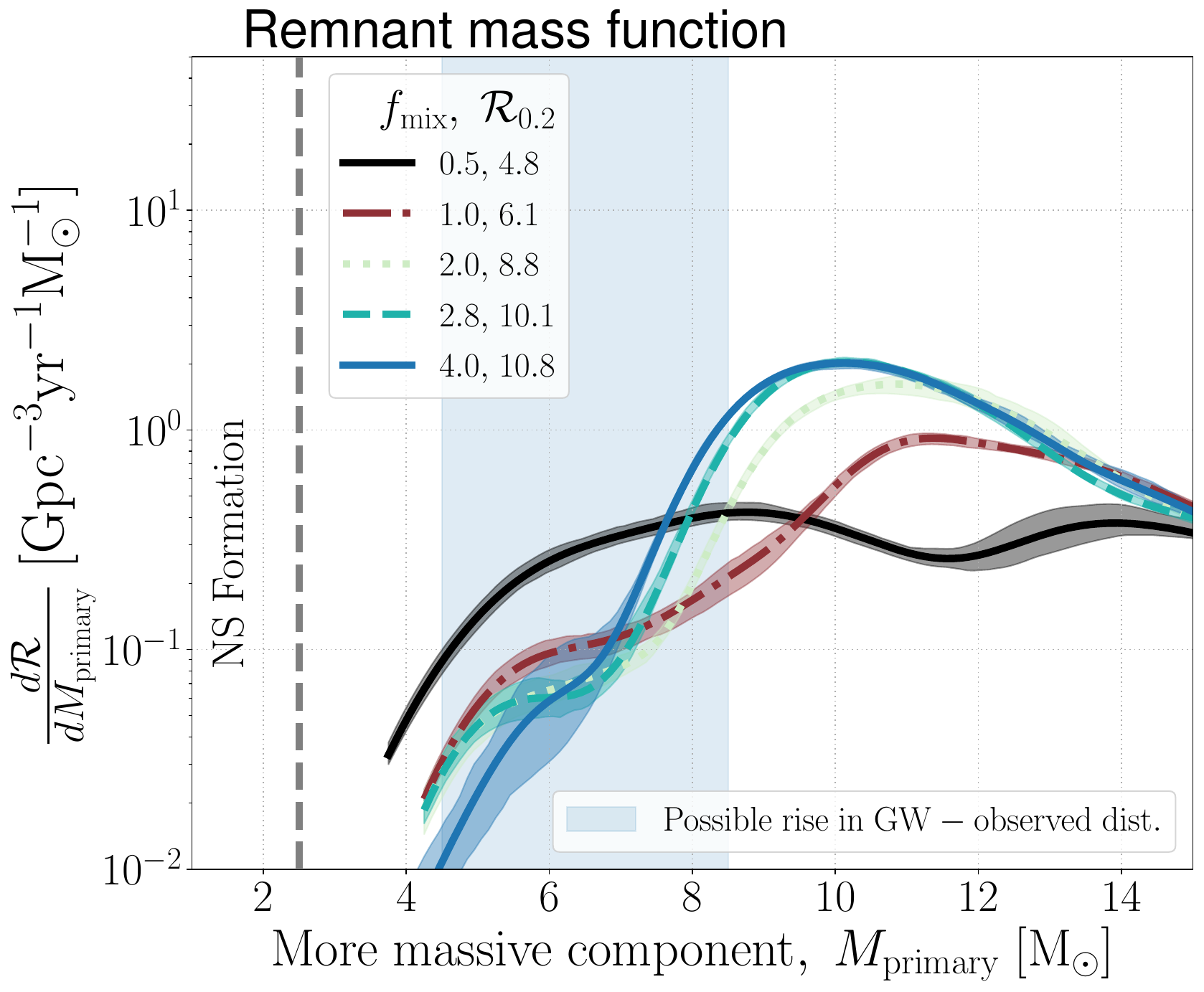}
\caption{Mass distributions in \Mmm for BBH and BHNS from the stable channel. We show variations in the supernova remnant mass function using the prescription from \cite{Fryer+2022}. Similar to the right column of Fig.~\ref{fig: validation and mass dists}, but using a kernel density distribution. To prevent the kernel to smooth over the cutoff mass in \Mmm, we only draw the distribution for \Mmm values where the corresponding histogram predicts a rate above $10^{-5}\Gpc^{-3}\yr^{-1}\Msun^{-1}$. 
\label{fig: dMsn} }
\end{figure}

We show the resulting \Mmm distribution of merging BBH and BHNS for the stable channel in Figure \ref{fig: dMsn}. All models predict the rate of systems with $\Mmm$ below about $4\Msun$ to be less than $10^{-5} \Gpc^{-3}\yr^{-1}$. In other words, all of these models predict a lack of BHs with masses below 4\Msun. This is not surprising since our fiducial model was chosen such that it is most efficient in forming low mass BHs. The variations in Figure \ref{fig: dMsn} are only expected to increase the gap between NS and BH masses.
We furthermore see that the overall merger rate density varies by a factor of about 2 between $\fmix = 0.5$ ($\Rlocal\approx5\Gpc^{-3}\yr^{-1}$) and $\fmix=4.0$ ($\Rlocal\approx11\Gpc^{-3}\yr^{-1}$).
Low $\fmix$ causes a shallow rise in the mass distribution with no clear peak. For higher values in \fmix, a peak starts to occur around 11\Msun. This peak becomes more pronounced and moves to lower \Mmm for increasing \fmix. For $\fmix=4.0$ the distribution peaks strongly at \Mmm~=~9.5\Msun, below which it decays steeply towards \Mmm~=~6\Msun.

The shape of the mass distribution is similar to the results from \cite{Olejak+2022} (top right panel of their Figure 5). In line with their results, we find the rate of \Mmm=6\Msun is much higher for \fmix =0.5 with respect to \fmix=4.0.
However, in contrast to \cite{Olejak+2022} we only show the contribution of the stable channel.
We speculate that this explains why the merger rate density between 3\Msun and 15\Msun is an order of magnitude higher in  \cite{Olejak+2022} with respect to our results.

\subsubsection*{Loss of orbital angular momentum through a circumbinary disk. \label{ss: angular momentum}}
A key ingredient determining the population of merging double compact objects is the orbital angular momentum loss during mass transfer that is not fully conservative. In order to form a binary compact enough to merge within a Hubble time through GW emission, it is generally crucial for the binary to shrink to a tight orbit during the second mass transfer phase. Which binaries manage to lose enough orbital angular momentum during this mass transfer phase will thus determine the shape of the mass distribution.

In our fiducial model we assume `isotropic re-emission' of matter during non-conservative stable mass transfer.
This means that mass lost from the donor star is assumed to be transported to the vicinity of the accretor (in the form of e.g.,\ an accretion disk), from where it is then ejected as a fast isotropic wind. 
Hence, the mass lost from the binary system carries the specific angular momentum of the accretor \citep[e.g.,][]{Soberman+1997}.
When mass is transferred at high rates, it is conceivable that some of the mass is lost through the L2 Lagrange point \citep[see e.g.,\ discussion in][]{Marchant2021}. This mass can end up in a circumbinary ring which removes angular momentum much faster than mass lost through isotropic-reemission \citep[e.g.,][]{ArtymowiczLubow1994,Soberman+1997,Renzo+2019,Lu+2022}.
An observational example of a system that has been argued to experience mass loss through L2 is SS433 \citep[][]{Fabrika1993,Fabrika2004}. One explanation of the observational outflow signatures of this system is mass loss through a circumbinary disc, see for example \cite{Cherepashchuk+2020} and references therein  \citep[for an alternative explanation to L2 mass loss see e.g.,][]{Blundell+2001}. 

\begin{figure*}
\centering
\includegraphics[trim=0.9cm 0.5cm 2.cm 2cm,clip=true,width=0.45\textwidth]{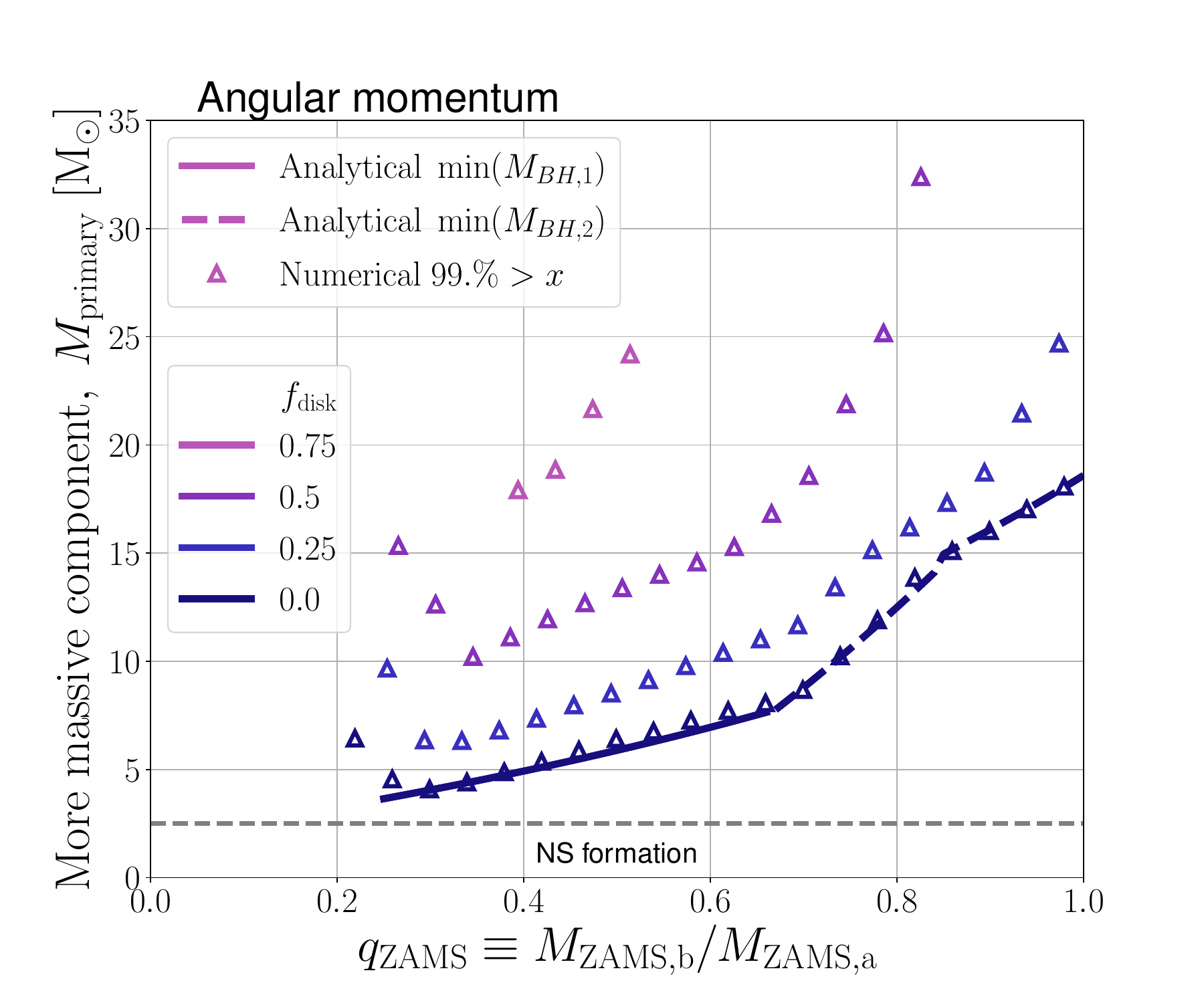}
\includegraphics[width=0.44\textwidth]{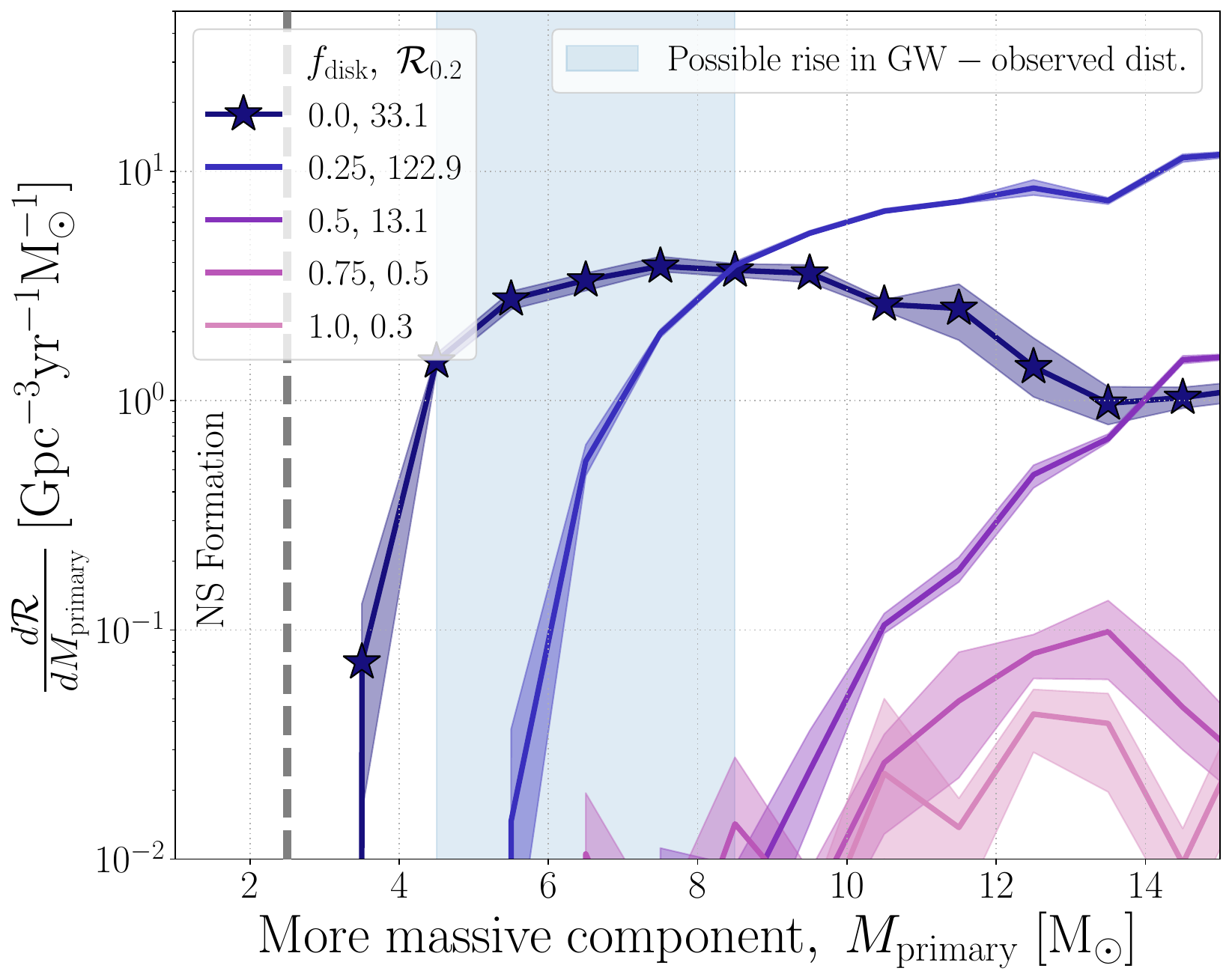}
\caption{Same as Figure \ref{fig: validation and mass dists}, but for variations in the fraction of the mass that is assumed to be lost from a circumbinary disk, \fdisk. {Because Equation \ref{eq: min MBH moreM} does not capture variations in the orbital angular momentum loss, we only show the analytical solution for $\fdisk=0.0$ in the left panel.} Furthermore, we do not show $\fdisk=1.0$ in the left panel because it contains too few samples to properly bin the distribution. 
\label{fig: gamma} }
\end{figure*}

We explore the effect of the specific angular momentum of mass lost from the system by assuming that a fraction \fdisk of the mass lost during every stable mass transfer event will be lost with the specific angular momentum of a circumbinary disk. 
We assume the circumbinary ring to be located at twice the orbital separation \citep[as first suggested by][]{TutukovYungelson1979}.
In Figure \ref{fig: gamma} we show variations of \fdisk ranging from $\fdisk=0$ (all mass is lost though isotropic-reemission, our fiducial model) to $\fdisk=1$ (all mass is lost from a circumbinary disk).

Variations in \fdisk have a significant impact on both the rate and the shape of the mass distribution (right panel of Figure \ref{fig: gamma}). Both the location and the peak of the mass distribution change.
Moreover, for \fdisk = 0.75 and 1.0, the stable mass transfer channel is effectively killed; the total local merger rate density is decreased to $0.5 \Gpc^{-3} \yr^{-1}$ and $0.3 \Gpc^{-3} \yr^{-1}$ respectively. We find this is mainly due to an increased number of stellar mergers. 
This result is in line with previous work that studied the effect of a circumbinary ring on the population of Be X-ray binaries and gravitational wave sources \citep[e.g.,][]{PortegiesZwart1995,DeDonderVanbeveren2004,MennekensVanbeveren2014,Vinciguerra+2020}.

Furthermore, the local merger rate density rises to about $141 \Gpc^{-3} \yr^{-1}$ for $\fdisk=0.25$. This is higher than the fiducial merger rate from the CE channel in \cite{vanson+2021}. For $\fdisk=0.5$ the rate has dropped back down to about $16 \Gpc^{-3} \yr^{-1}$, which implies that the contribution of the stable mass transfer channel experiences some maximum in the local merger rate density between $\fdisk=0$ and $\fdisk=0.5$.
The actual value of \fdisk most likely depends on the mass transfer rate \citep[see ][ for a detailed analysis]{Lu+2022}. \cite{Lu+2022} find that for non-extreme mass ratios (not much less, or much greater than one), \fdisk can become of order unity for rates $\gtrsim \rm{few} \times 10^{-4}\Msun\yr^{-1}$.

Finally, the minimum primary mass from the stable channel increases as a larger fraction of the mass is lost though a circumbinary disk. In other words, higher \fdisk correspond to a higher value of \minmbh at constant \qzams.
This can be seen in the left hand panel of Figure \ref{fig: gamma}.
This figure also shows that \minmbh increases with \qzams, following a similar trend as that described by Equations \ref{eq: minMBH1}, \ref{eq: minMBH2} and \ref{eq: min MBH moreM}.

\section{Discussion \label{sec: discussion} }
In this work, we investigate the low-mass end of the primary mass distribution ($\Mmm$) for BBHs and BHNS systems as predicted from the stable mass transfer channel.
We find that the stable mass transfer channel leads to a sharp cut-off at the low-mass end of the primary mass distribution. This feature is a consequence of the requirement of stable mass transfer, which is a characteristic property of the channel.
We analytically express the minimum allowed primary mass, \minmbh, as a function of the ZAMS mass ratio \qzams. We identify the key physical processes that determine the value of \minmbh, and discuss the robustness of this minimum against variations. Depending on the adopted physics, we find that \minmbh leads to a low-mass cut-off in the primary masses between $2.5-9\Msun$.
Our main results as presented in Figure \ref{fig: validation and mass dists} provide several direct predictions.

\subsection{Remnant mass function or binary physics effect?}
Many of the physics variations explored in this work lead to a dearth of merging BBH and BHNS systems with low primary masses. 
This lack of low-mass BHs also dictates the location of the peak of the BBH primary mass distribution. In this case, the shape of the mass distribution at the low-mass end is thus determined by binary physics. 

Alternatively, adopting a remnant mass function with a low-mass gap can also cause the models to predict a pile up just above the upper edge of this gap.
Several isolated binary evolution models predict a peak near 10\Msun when adopting the `RAPID' SN engine prescription from \cite{Fryer+2012} \citep[see e.g.,][]{Belczynski+2016Natur,GiacobboMapelli2018,Giacobbo+2018,Wiktorowicz+2019,Belczynski+2020,Tanikawa+2022}.
In this case, the remnant mass function determines the shape of the low mass end of the mass distribution.

The crucial difference between these two scenarios is that the remnant mass distribution is expected to affect \textit{all BH and NS formation}, while we expect the constraints discussed in this work to affect \textit{only those systems that evolve through the stable channel} i.e., that have experienced two phases of stable mass transfer.

A smoking gun to determine whether the stable mass transfer channel dominates the low-mass end of the BH mass distribution observed in GW would thus be if the dearth of low-mass BHs persists in the distribution of primary BH masses observed in GW, while a significant number of low-mass BHs are detected as part of systems that are not expected to have evolved through the stable mass transfer channel. Examples of the latter are low-mass XRB (see the discussion on XRB below in Section \ref{sss: xrb})

\subsection{NSNS and binary white dwarf mergers}
In principle, the arguments presented in this work should hold for all binary systems that have experienced stable mass transfer from the initially more massive to the initially less massive star, and vice versa.
This implies that the stable mass transfer channel is inefficient at producing lower mass systems like NSNS.
This finding agrees with earlier work that suggests the formation of NSNS mergers is dominated by the CE channel \citep[e.g.,][]{vignaGomez+2018,Chruslinska+2018_NSNS}.
Earlier work has also found that different channels dominate the formation of NSNS mergers with respect to BBH mergers (see the appendix of \citealt{Wagg+2021} and the discussion in  \citealt{Broekgaarden+2021b}).
If we assume that the CE channel dominates the formation of NSNS systems, while the stable mass transfer channel dominates the shape of the primary mass distribution around the peak at $9\Msun$, then the transition between these two channels happens within a narrow range of remnant masses.
This would have large implications on the efficiency of CE for different donor masses, as it suggests that successful CE ejection is only possible for lower mass stars that produce NS \citep[see also][]{Klencki+2020,Klencki2021}.

A similar constraint on the primary mass could be explored for binary White Dwarf (WD) formation, however, many of our assumptions (such as supernova mass loss, and an approximately constant core mass fraction $\fcore\approx0.34$, see also Appendix \ref{app: alt minimum mass}) cannot simply be directly adopted for WD progenitors.
In the context of the formation of double WDs, \cite{Woods+2012} emphasized the importance of systems in which the first phase of mass transfer is stable but the second mass transfer phase is unstable. Numerous works investigate formation channels in which the last mass transfer phase  (which resulted in the double WD) is stable \citep[e.g.,][]{Nelson+2004,Kalomeni+2016,SunArras2018,Chen+2022}.
An analysis similar to the one in this paper might be used to study the potential population of double WDs formed following only stable mass transfer in both directions.

\subsection{Results in context of X-ray binary observations \label{sss: xrb} }
In this work, we have discussed a potential dearth of BHs with low masses as observed in GW events.
The original proposal for a gap in the mass distribution between NS and BHs was based on the detection of X-ray binaries \citep[XRB, ][]{Bailyn+1998,Ozel+2010,Farr+2011}. One might therefore wonder if the stability criteria discussed in this work could also lead to a dearth of low-mass BHs in observed XRBs. However, it is unclear whether XRB systems and GW progenitors belong to the same astrophysical population \citep[see e.g. ][]{FishbachKalogera2021,Belczynski+2021}. It is difficult to resolve this issue because the observed population of XRB represents a wide variety of binary star evolutionary stages.
In order to understand our results in context of XRB observation, we take a closer look at the XRB populations that were used to infer a NS-BH mass gap in the first place.

The population of XRB is commonly subdivided into two classes, characterised by the mass of the donor star.
First, there are low-mass XRB, where the compact objects accretes from low-mass donor stars below about 2–3\Msun through Roche-lobe overflow.
The origin of short period low-mass XRB is unknown, but it is most commonly assumed that they are the outcome of a CE event \citep[e.g.,][for a discussion on plausible evolutionary origins]{Podsiadlowski+2003}. However, many different evolutionary pathways have been proposed \citep[e.g. ][]{EggletonVerbunt1986,Ivanova2006,MichaelyPerets2016,Klencki+2017}.
Due to the extreme ZAMS mass ratios required to form a compact object + a low-mass companion, we \textit{do not} expect the first mass transfer phase to be stable, and thus we do not expect the stable mass transfer channel to contribute to the population of low-mass XRB.
Hence, if there is truly a dearth of low-mass BHs in low-mass XRB, this would not be caused by the stability requirements discussed in this work.

Secondly, there are high-mass XRB, which accrete from a typically higher-mass ($\gtrsim5\Msun$) companion star. Due to the longer timescales involved, these systems are often expected to be wind fed as opposed to experiencing stable RLOF (possibly occurs in phase D of Figure \ref{fig: cartoon}).
In this work, we have found that the stability of the \textit{second} mass transfer phase is is a crucial element in \minmbh. Hence we also do not expect the mechanisms as discussed in this work to lead to any dearth of low-mass BHs in high-mass XRB.

There is third population of XRB systems; Wolf-Rayet X-ray binaries (or WR-XRB in short), which are expected to be the direct descendants of high-mass XRB. They are composed of a (stripped) helium star and a compact object and exist on the He burning nuclear timescale. One would thus expect the birthrate of WR-XRB systems to be approximately equal to the birthrate of high-mass XRB. However, while there are hundreds of Galactic high-mass XRB \citep[see][for the most recent review]{HMRXB_Liu+2006}, there is only one known WR-XRB system in the Milky Way \citep[Cygnus X-3][]{CygX3_1992+vanKerkwijk}. This is known as the `missing WR-XRB’ problem \citep[e.g.,][]{Lommen+2005}.
\cite{van-den-Heuvel+2017} argue that this problem can be explained based on arguments of mass transfer stability in the same way as we explain a lack of low-mass BHs in the population of GW sources: only when the mass ratio at the second mass transfer phase is in the right regime for stable mass transfer, can the system avoid CE evolution.
Although the results in our work do not explain a dearth in low or high-mass XRB, they \textit{can} provide an explanation for the missing WR-XRB problem as well as an explanation for a dearth of  primary BHs with low masses, inferred from GW events.

As mentioned above, if the dearth of low-mass BHs persists in the distribution of primary BH masses observed in GW while a significant number of low-mass BHs are detected as the less massive components of GW events or as part of low- and high-mass XRB, this could serve as a smoking gun to determine whether the stable mass transfer channel dominates the low-mass end of the mass distribution observed in GW.
On the other hand, if a dearth of low-mass BHs remains in \textit{all} mass-observations of BHs, we argue that a gap in the remnant mass distribution is a more likely explanation.
A rapidly increasing number of recent detections through various observational methods already seem to challenge whether the NS-BH mass gap is empty \citep[e.g.,][]{Thompson+2019_gapdetection,Giesers+2019,Breivik+2019,Rivinius+2020,WyrzykowskiMandel2020,GomezGrindlay2021,Sahu+2022,Lam+2022,vanderMeij+2021,Jayasinghe+2021,Jayasinghe+2022,Andrews+2022}. At the same time, many of these candidates are controversial \citep[see][and references therein]{ElBadry+2022}, and the existence of a gap in the remnant mass distribution remains an open question to this day.
A large increase in BH mass measurements is expected from both GW observations \citep[][]{LVK_livingReview2018}, as well as from detections of BH + main sequence systems in the Gaia data release 3 \citep[e.g.,][]{Breivik+2017,MashianLoeb2017,Andrews+2019,Langer+2020,Andrews+2021,Chawla+2022,Janssens+2022,GaiaDR3_AstrometricBinaries_2022}. 
Hence we are hopeful that near future detection surveys will provide evidence in favour or against the existence of a NS-BH mass gap.

\subsection{Filling the low-mass gap from below \label{ss: filling lowM gap from below} }
Several works investigated if it is possible to populate the lower mass gap between 3-5\Msun through hierarchical mergers.
\cite{SamsingHotokezaka2021} considered NSNS merger products in dense cluster environments. They conclude that populating the low-mass gap through in-cluster mergers of NSs is a much too slow process to be relevant, even for a highly idealised case.
In response to the detection of GW190814 \citep[a compact binary coalescence involving a less massive component with a mass of 2.50-2.67\Msun][]{GW190814_detection2020}, \cite{LuBeniaminiBonnerot2021} propose that GW190814 was a second-generation merger from a hierarchical triple system.
They anticipate that this scenario would lead to a narrow peak in the mass distribution of the less massive component masses between 2.5 and 3.5\Msun. They find that it is plausible, but rare for a NSNS merger to give rise to a second-generation merger and estimate that 0.1 to 1 per cent of NSNS mergers occurring in triples could contribute to this channel.
Similarly \cite{Hamers+2021} consider repeated mergers of NSs and BHs in stellar 2+2 quadruple systems and find that second generation mergers are about ten million times less common that first generation counterparts.
Hence we do not expect hierarchical mergers to `fill the gap from below', nor cause a peak at about 9\Msun.

\subsection{Caveats \label{ss: disc. caveats}}

\paragraph{Adopting a fixed value for the accretion efficiency}
In the model variations presented in Section \ref{sec: validation and behaviour}, we have adopted a fixed value for \betaacc = 0.5 \citep{MeursVandenHeuvel1989,BelczynskiStarTrack+2008,Dominik+2012} .
In contrast, in the model shown in Figure \ref{fig: fiducial mass dist} we adopt an accretion rate that is limited to the thermal timescale of the accretor to simulate accretors that remain in thermal equilibrium. This limits the accretion rate to $\dot{M}_{a} = C \times M_a/t_{KH,a}$, where $M_a$ and $t_{KH,a}$ are the mass and Kelvin-Helmholz time of the accretor, and C=10 is a constant factor assumed to take into account the expansion of the accreting star due to mass transfer \citep{PaczynskiSienkiewicz1972,Neo+1977,Hurley+2002,Schneider+2015}.
Adopting this accretion rate will cause \betaacc to be effectively zero for binary systems with low $\qzams\sim0.3$ \citep[see e.g. the top panels of Figures 19 and 20 from][]{Schneider+2015}.
The value of \minmbh is lowest at low values of \betaacc and \qzams (Figure \ref{fig: validation and mass dists}), and such systems will thus pollute any dearth in the mass range $\Mmm = 2.5-6\Msun$.

It is hard to say what what the real accretion rate will be, since this depends critically on the response of the accretor which is here merely encompassed in the constant $C$.
A more realistic treatment of the expanding accretor could also affect mass-transfer stability, since this expansion may lead to a contact phase and subsequent CE evolution \citep[see e.g.,][]{Pols1994,LangerHeger1998,Justham+2014}. On top of this, the post-mass-transfer properties of the accreting star are not captured by single-star models \citep{RenzoGotberg2021}, and will further influence the details of the second mass-transfer phase \citep{Renzo+2022}.

\paragraph{Treatment of Case A mass transfer}
Mass transfer where the donor star overflows its Roche lobe while still on the main sequence is known as `Case A' mass transfer.
In general, rapid population synthesis simulations oversimplify the processes involved in a mass transfer episode, but the outcome of Case A mass transfer is particularly difficult to predict \citep[e.g.,][]{Pols1994,Sen+2022}.
In this work, we adopt a set value of $\zeta_{MS} = 2$ to determine the stability of mass transfer for donor stars on the main sequence (see Section \ref{sec: definitions}). In our simulations, Case A mass transfer is thus more prone to unstable mass transfer, which in part explains why we find that Case A mass transfer is subdominant in the stable mass transfer channel.
Our simulations under-predict the size of the donor's core following case-A mass transfer. Nonetheless Case A mass transfer is generally assumed to lead to smaller core masses and to be more conservative than Case B mass transfer \citep[e.g.,][]{Schneider+2015,Sen+2022}, and we expect the former prediction to hold even when core masses are corrected since both smaller cores and more conservative mass transfer lead to higher values of \minmbh (see Figure \ref{fig: validation and mass dists}). Hence we find that systems from Case A mass transfer are not dominant in determining the cut-off mass in \Mmm.

\section{Conclusions \label{sec: conclusions}}
We explore the low-mass end of the primary mass distribution of BBH and BHNS systems that can lead to GW sources. We argue that a dearth of BHs with masses between 3-5\Msun, as observed in the GW-inferred mass distribution, should be jointly investigated with the observed peak of primary masses at about 9\Msun.
With this in mind, we investigate the stable mass transfer channel to GW emitters. We make predictions for the expected merger rates and mass distributions that follow from this channel, and explain their origins. Our main findings are listed below:

\begin{enumerate}

\item  \textit{The low mass end of the primary BH mass distribution inferred from GW detections can be explained remarkably well by the stable mass transfer channel alone.}
    For our fiducial assumptions, we naturally match the local rate ($20\Gpc^{-3}\yr^{-1}$ at redshift $0.2$) and key features of BBH mass distribution (the dearth of primary masses between 2.5--6\Msun, and the subsequent peak around 8--10\Msun) without need for additional channels (see Figure~\ref{fig: fiducial mass dist}). 

\item \textit{ A unique prediction of the stable channel is that it is unable to produce GW events with primary BH masses below a certain cut-off mass.}
The reason for the existence of the cut-off is (1) the requirement of stability during the mass transfer phases, which imposes constraints on the mass ratios, and (2) the fact that the final BH masses do not simply scale with the initial mass.
Specifically, at the onset of the second mass transfer phase, the masses of the binary components can be expressed as a function of the initial masses.  This places a bound on the zero-age mass of the initially most massive star and consequently the mass of the BH it gives rise to. Similarly, the requirement of stability during the second mass transfer phase places bounds on the mass of the compact object resulting from secondary star (see Section~\ref{sec: definitions}).

\item \textit{Our results imply that the binary physics involved in the stable channel alone can provide an explanation for the purported NS-BH mass gap in GW detections.}
This is an alternative explanation to the common assertion that the gap results from  supernova physics.
This also implies that GW detections may not directly reflect the remnant mass function, as selection effects of the formation channels can not be neglected.

\item \textit{We provide an analytical expression for the lower limit for the cut-off mass }
    We find expressions for the binary components at all relevant stages using parameterised assumptions for the dominant physical processes (see Figure \ref{fig: cartoon}), namely, the mass transfer efficiency, the core mass fraction, the mass transfer stability and the difference between the core mass and final remnant mass (Equations \ref{eq: minMBH1}, \ref{eq: minMBH2} and \ref{eq: min MBH moreM}).

\item \textit{Using numerical simulations, we conduct an extensive exploration of the uncertain physical processes that impact the stable channel.} We show these impact the shape of the low end of the mass distribution and location of the peak. (Figure~\ref{fig: validation and mass dists}, \ref{fig: dMsn} and \ref{fig: gamma}).

\item \textit{The difference between the remnant mass function inferred from electromagnetic observations and the mass distribution from GW observations may serve as a smoking gun.} Specifically, if the NS-BH gap fills in for electromagnetic observations but remains for GW observations, this would be a telltale sign of a dominant contribution by the stable channel in this mass range.

\end{enumerate}

\begin{acknowledgments}
LvS would like to thank Eric Burns, Chris Fryer and David Hendriks useful comments and informative discussions on the impact of the supernova remnant mass prescriptions. 
LvS performed portions of this study as part of the pre-doctoral Program at the Center for Computational Astrophysics of the Flatiron Institute, supported by the Simons Foundation.
The authors thank the anonymous reviewer for helpful comments that improved this work. 
The authors acknowledge partial financial support from the  National Science Foundation under Grant No. (NSF grant number 2009131  and PHY-1748958), the Netherlands Organisation for Scientific Research (NWO) as part of the Vidi research program BinWaves with project number 639.042.728 and the European Union’s Horizon 2020 research and innovation program from the European Research Council (ERC, Grant agreement No. 715063). EZ acknowledges funding support from the European Research Council (ERC) under the European Union’s Horizon 2020 research and innovation programme (Grant agreement No. 772086).
\end{acknowledgments}


\section*{Software and data\label{sec:cite}} 
{All code associated to reproduce the data and plots in this paper is publicly available at \url{https://github.com/LiekeVanSon/LowMBH_and_StableChannel}.
The data used in this work is available on Zenodo under an open-source Creative Commons Attribution license at\dataset[10.5281/zenodo.7080725]{https://zenodo.org/record/7080725}, and  \dataset[10.5281/zenodo.7080164]{https://zenodo.org/record/7080164}.
Simulations in this paper made use of the \COMPAS rapid binary population synthesis code (v02.26.03), which is freely available at \url{http://github.com/TeamCOMPAS/COMPAS}  \citep{COMPAS_method}.
This research has made use of GW data provided by the GW Open Science Center (\url{https://www.gw-openscience.org/}), a service of LIGO Laboratory, the LIGO Scientific Collaboration and the Virgo Collaboration.}
Further software used:
astropy \citep{2013A&A...558A..33A,2018AJ....156..123A},  Python \citep{PythonReferenceManual},  Matplotlib \citep{2007CSE.....9...90H},  {NumPy} \citep{2020NumPy-Array}, SciPy \citep{2020SciPy-NMeth}, \texttt{ipython$/$jupyter} \citep{2007CSE.....9c..21P, Kluyver2016jupyter},  Seaborn \citep{waskom2020seaborn}  and  {hdf5}   \citep{collette_python_hdf5_2019}.

\appendix

\section{The dependence of mass transfer stability on the mass ratio and the mass transfer accretion fraction \label{app: zeta} }
\begin{figure}
\centering
\includegraphics[width=0.55\textwidth]{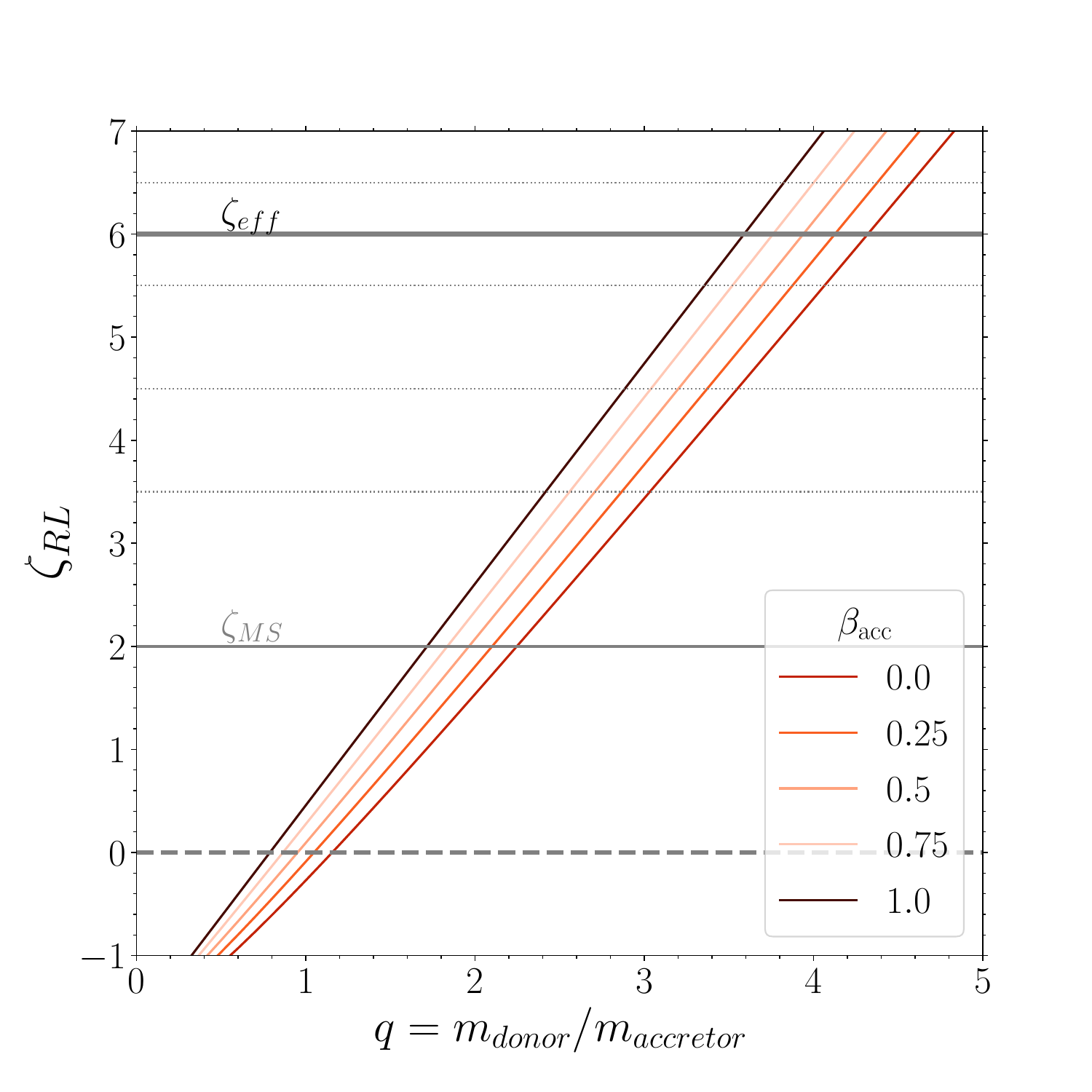}
\caption{$\zeta_{\rm{RL}}$ as a function of mass ratio of between the donor and accretor star. Mass transfer is dynamically stable as long as $\zeta_{\rm{RL}} \leq \zetaeff$. The intersections of $\zeta_{\rm{RL}}$ and \zetaeff reveal different values of \qcritI and \qcritII.
Our default value of $\zetaeff=6.0$ for a star with a clear core-envelope structure is anotated. $\zeta_{\rm{MS}} = 2$ shows the adopted stability criteria assumed for main-sequence stars \protect\citep{Ge+2015} \label{fig: zetas}}
\end{figure}

In Figure \ref{fig: zetas} we show $\zeta_{\rm{RL}}$ as a function of \qzams. \footnote{The full functional form of $\zeta_{\rm{RL}}$ can be found at \url{https://github.com/LiekeVanSon/LowMBH_and_StableChannel/blob/master/Code/AppendixFig6_zeta_q_beta_relations.ipynb}, where we closely follow \cite{Soberman+1997}. } 
Mass transfer is dynamically stable as long as $\zeta_{\rm{RL}} \leq \zetaeff$.
The intersection of the coloured lines with the adopted value of \zetaeff (grey horizontal lines) lead to a value of $q_{\rm{crit}}$. For example, if we assume $\betaacc = 0.0$ for the second mass transfer, and $\zetaeff=6.0$, we can see $\qcritII = m_{\rm{donor}}/m_{\rm{accretor}} \approx 4.4$ for this mass transfer phase.
Note that we define $\qcritI = \MZAMSII/\MZAMSI $ which is the inverse of $m_{\rm{donor}}/m_{\rm{accretor}}$.

\section{Mass dependent core mass fraction \label{app: alt minimum mass} }
Throughout this work we have assumed that the difference in mass between the core post mass transfer and the final remnant mass is nonzero (i.e., $\dMsn \neq 0$).
Note that we use $\dMsn$ as a shorthand for \textit{all} mass lost between the core mass post mass transfer and the final remnant mass, i.e.\ including stellar winds such as Wolf-Rayet-like winds.
We have adopted this because we find that this leads to a more stringent constraint on the BH and NS masses that form from the stable channel.
However, in some cases such as the formation of double WDs through stable (early) Case B mass transfer, assuming $\dMsn = 0$ may be closer to the truth.

In this section we thus look at an alternative to  Equation \ref{eq: minMzams dep Msn}, by assuming $\dMsn=0$, but \fcore is a function of the ZAMS mass;

\begin{equation}
    f_{core1} = a_f M_{\mathrm{TAMS} } + b_f ,
    \label{eq: fcore}
\end{equation}
where $M_{TAMS}$ refers to the mass at the terminal age main sequence (TAMS). We approximate $M_{\mathrm{TAMS,1}}=\MZAMSI$ and $M_{\mathrm{TAMS,2}}=\MtildeII$.
Applying this to  Equation \ref{eq: rewrite}, we get
\begin{equation}
    \frac{\qzams + \betaacc}{\qcritII + \betaacc} \leq \fcore = a_f \MZAMSI + b_f
    \label{eq: method 2}
\end{equation}

Note that we define all mass ratios (including \qcritI and \qcritII)  as the ratio between the initially less massive component over the initially more massive component.
This means that for the first mass transfer phase, mass transfer will be dynamically stable as long as $\MZAMSII/\MZAMSI = \qzams = M_{\rm accretor}/M_{\rm donor} \geq \qcritI$. While for the second mass transfer $\qpremtII =  M_{\rm b}/M_{\rm a} = M_{\rm{post MT1}}/\MbhA = M_{\rm donor}/M_{\rm accretor} \leq \qcritII$.

And thus
\begin{equation}
    \MZAMSI \geq \frac{1}{a_f} \left[ \frac{\qzams + \betaacc}{\qcritII + \betaacc} - b_f \right]
\end{equation}

The minimum cut-off mass is reached for $\qzams = \qcritI$, which leads to:

\begin{equation}
\boxed{
    \MZAMSI \geq \frac{1}{a_f} \left[ \frac{\qcritI + \betaacc}{\qcritII + \betaacc} - b_f \right]
    }
\label{eq: minMzams dep fc}
\end{equation}

Applying this to equations \ref{eq: minMBH1}, \ref{eq: minMBH2} we get a different relation for \minmbh from  Equation \ref{eq: min MBH moreM}.

\begin{table}
\begin{center}
\caption{Merger rates of BBH, BHNS and their combined rate at redshift 0.2, for the core mass fraction, mass transfer stability and mass transfer efficiency variations of the stable channel (as described in Section \ref{ss: validation}). \label{tab: rates1}}
\begin{tabular}{l | lllll | lllll | lllll}
 [$\Gpc^{-3} \yr^{-1}$] & \multicolumn{5}{c}{Core mass fraction   ($\fcore$)  } & \multicolumn{5}{c}{Mass transfer stability ($\zetaeff$)} & \multicolumn{5}{c}{Mass transfer   efficiency ($\betaacc$)} \\  \hline \hline
Variations           & 0.27     & 0.31     & 0.34    & 0.374    & 0.408    & 3.5      & 4.5      & 5.5       & 6        & 6.5      & 0         & 0.25      & 0.5       & 0.75      & 1        \\ \hline
$\mathcal{R}_{\rm BHNS, 0.2}$ & 0.7      & 2.7      & 7.4     & 11.7     & 15.5     & 0        & 0.3      & 3.3       & 6.5      & 13       & 3.9       & 6.6       & 6.5       & 0.1       & 0        \\
$\mathcal{R}_{\rm BBH, 0.2}$  & 3.5      & 10       & 25.8    & 53.9     & 87.3     & 0.9      & 4.6      & 16.7      & 25.3     & 35.9     & 10.1      & 17.4      & 25.3      & 34        & 39.6     \\
$\mathcal{R}_{0.2}$  & 4.2      & 12.7     & 33.2    & 65.5     & 102.8    & 0.9      & 4.8      & 19.9      & 31.7     & 48.9     & 14        & 24        & 31.7      & 34.1      & 39.6    
\end{tabular}
%
\centering
\caption{Merger rates of BBH, BHNS and their combined rate at redshift 0.2, for the supernova prescription and angular momentum variations of the stable channel (as described in Section \ref{ss: dMsn and gamma}).  \label{tab: rates2}}
\begin{tabular}{l | lllllll | lllll}
[$\Gpc^{-3} \yr^{-1}$] & \multicolumn{7}{c}{Supernova prescription   ($f_{\rm mix}$)} & \multicolumn{5}{c}{Angular momentum   ($f_{\rm disk}$)} \\ \hline \hline
Variations           & 0.5    & 0.7    & 1      & 1.4    & 2      & 2.8    & 4      & 0         & 0.25       & 0.5       & 0.75     & 1       \\ \hline 
$\mathcal{R}_{\rm BHNS, 0.2}$ & 0.6    & 0.6    & 0.7    & 0.9    & 0.8    & 0.9    & 1.1    & 7.1       & 4.5        & 1.1       & 0.4      & 0.3     \\
$\mathcal{R}_{\rm BBH, 0.2}$  & 4.2    & 4.1    & 5.5    & 6.8    & 8      & 9.2    & 9.7    & 26        & 118.4      & 12        & 0.1      & 0       \\
$\mathcal{R}_{0.2}$  & 4.8    & 4.7    & 6.1    & 7.8    & 8.8    & 10.1   & 10.8   & 33.1      & 122.9      & 13.1      & 0.5      & 0.3    
\end{tabular}
\end{center}
\end{table}

\section{Chirp mass and final mass ratios \label{app: other mass dists}}
\begin{figure}

\includegraphics[width=0.49\textwidth]{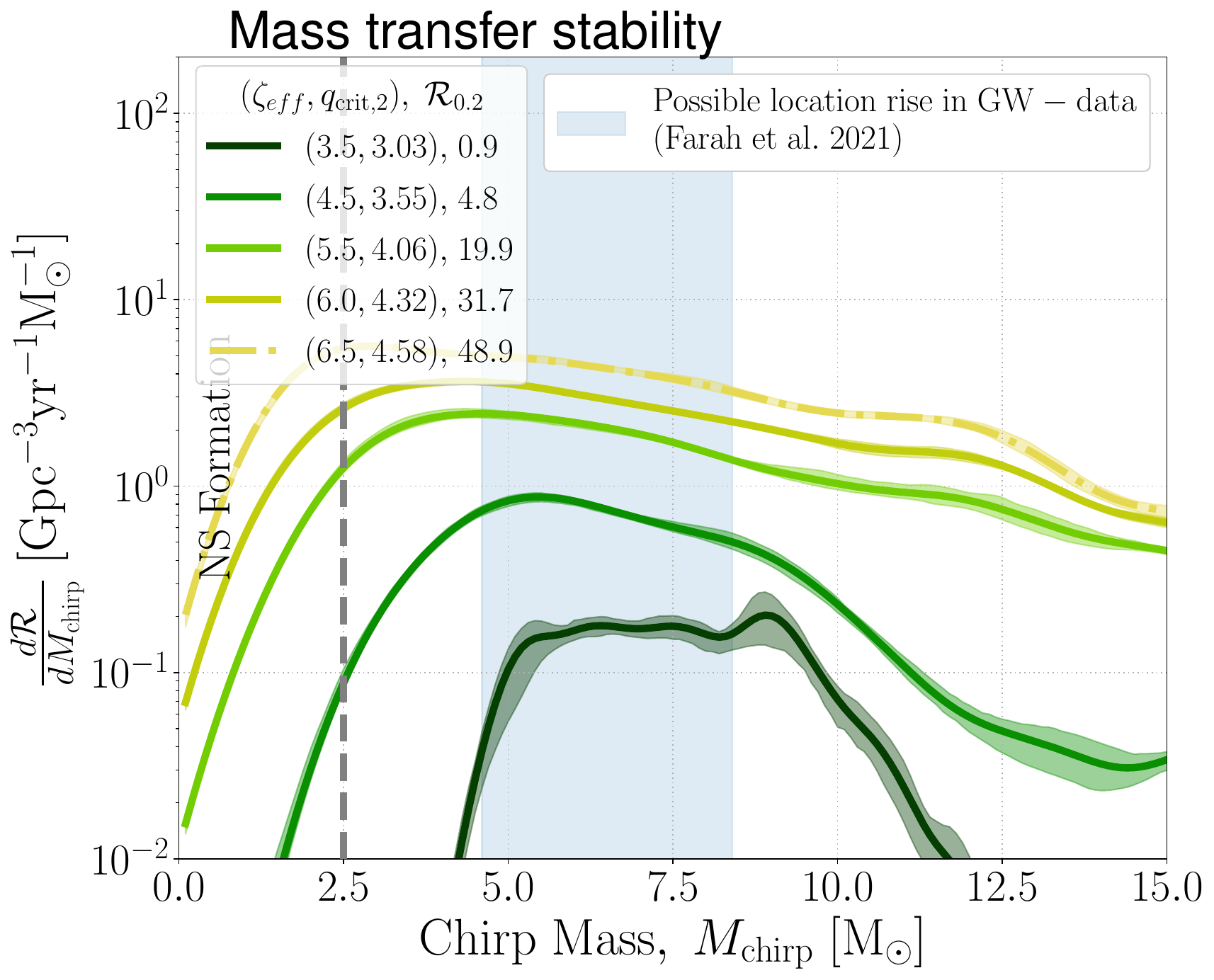}
\includegraphics[width=0.49\textwidth]{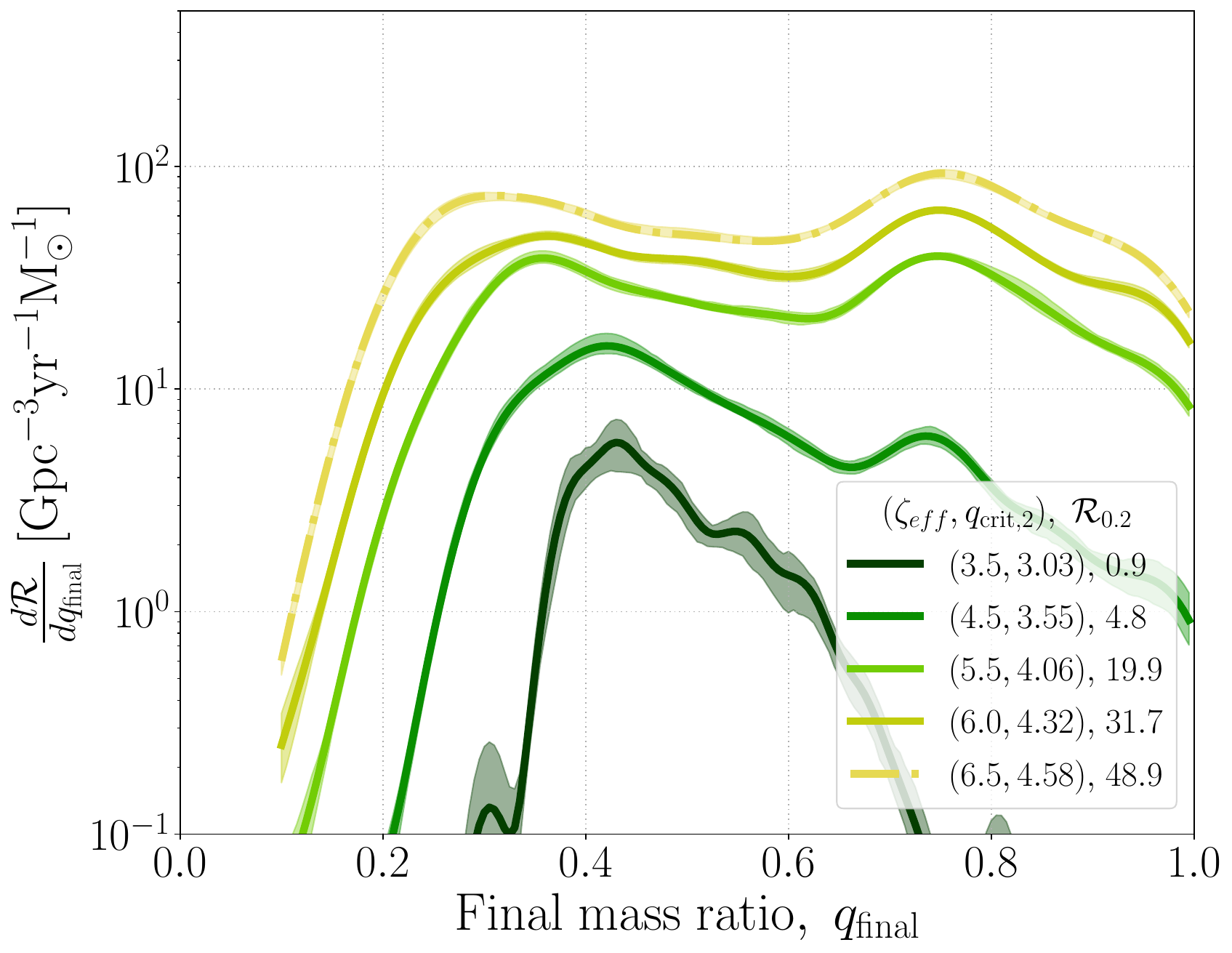}

\includegraphics[width=0.49\textwidth]{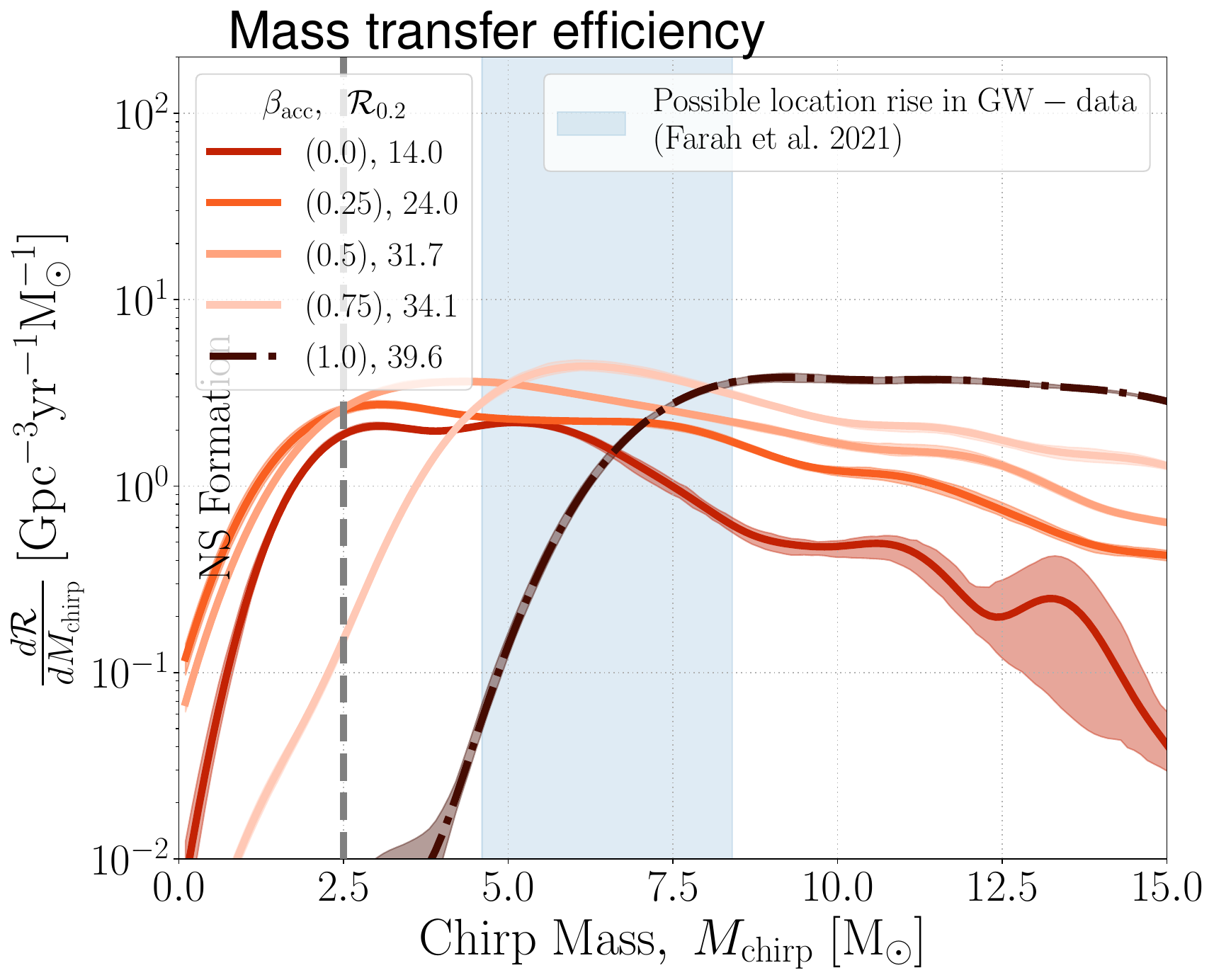}
\includegraphics[width=0.49\textwidth]{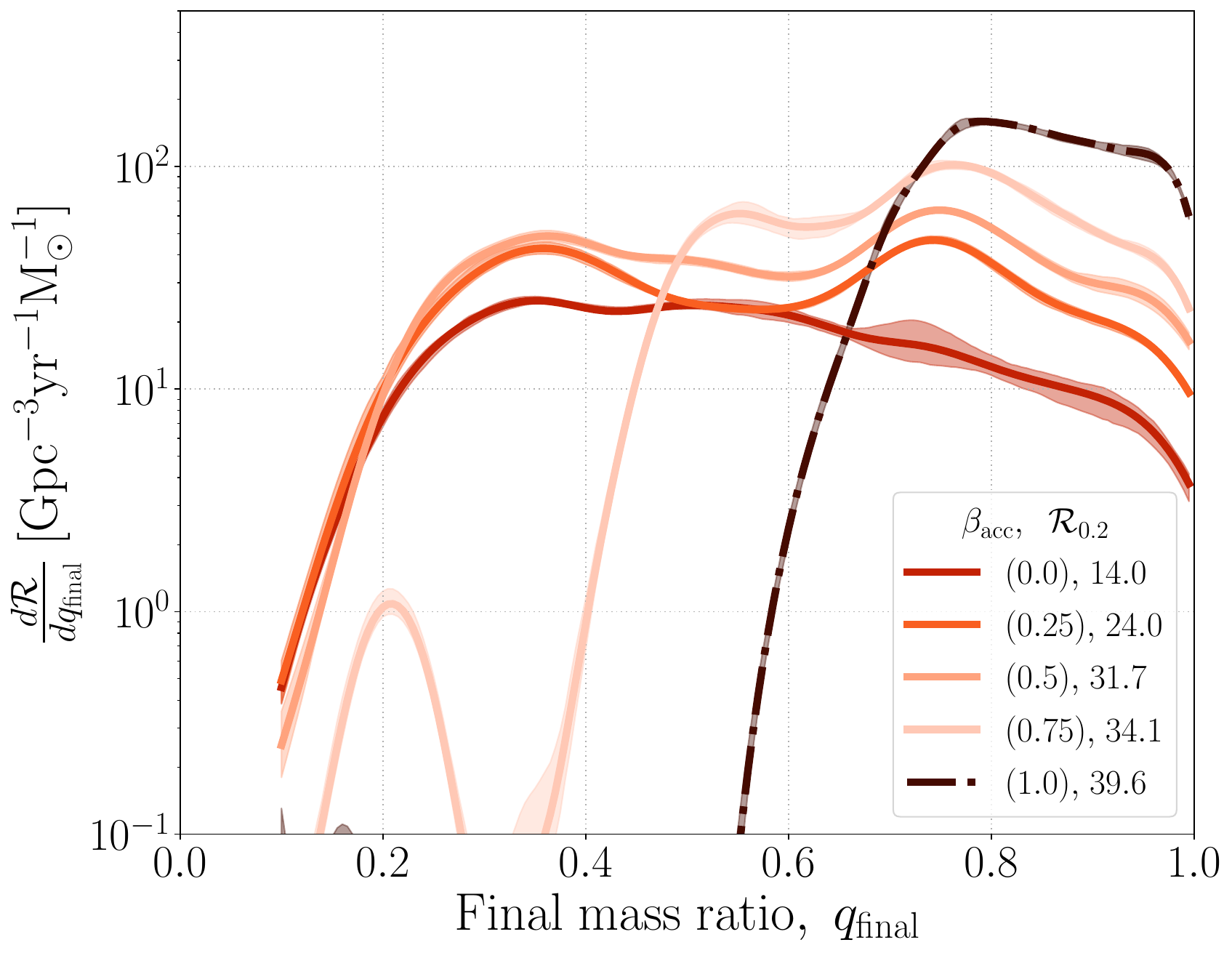}

\includegraphics[width=0.49\textwidth]{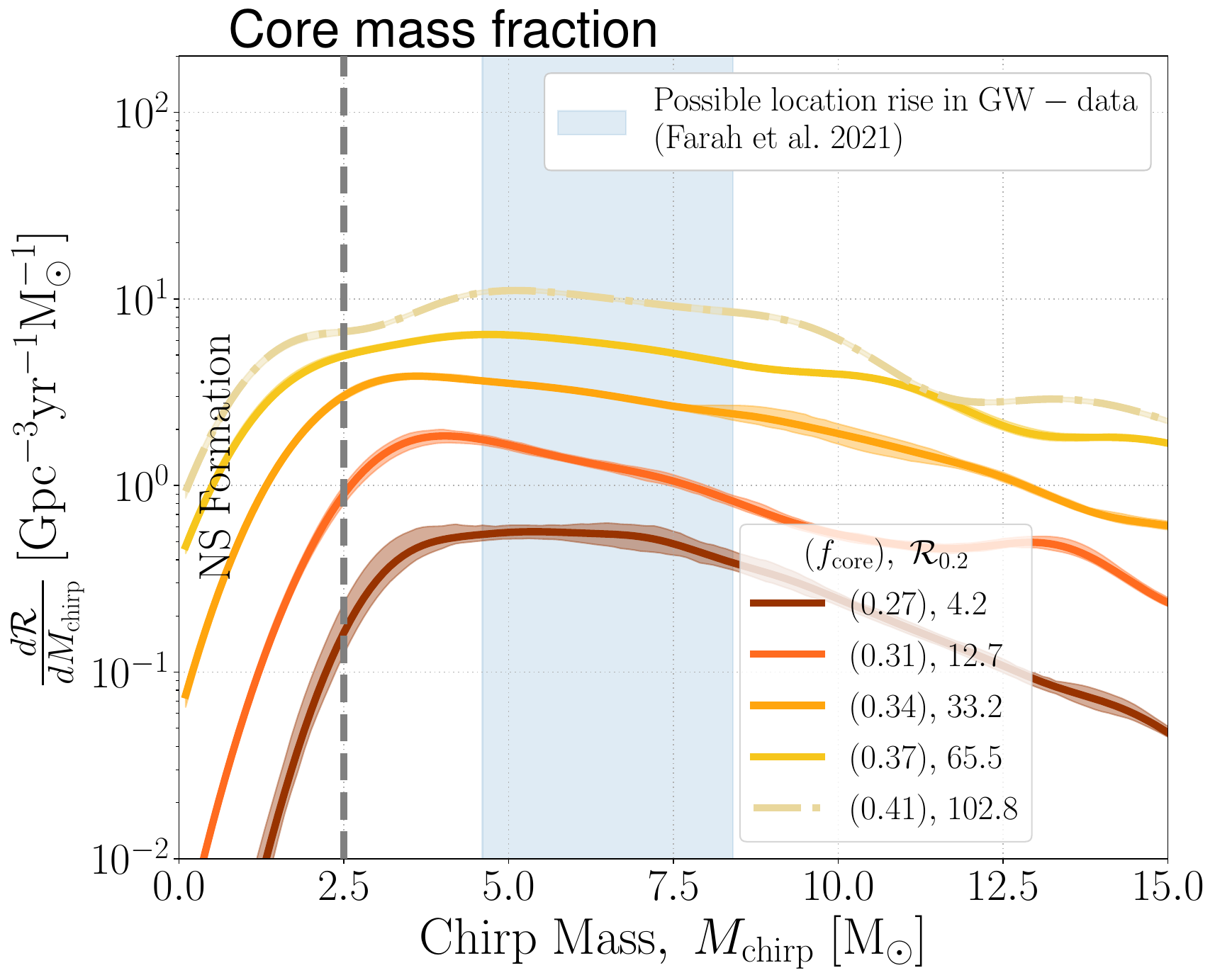}
\includegraphics[width=0.49\textwidth]{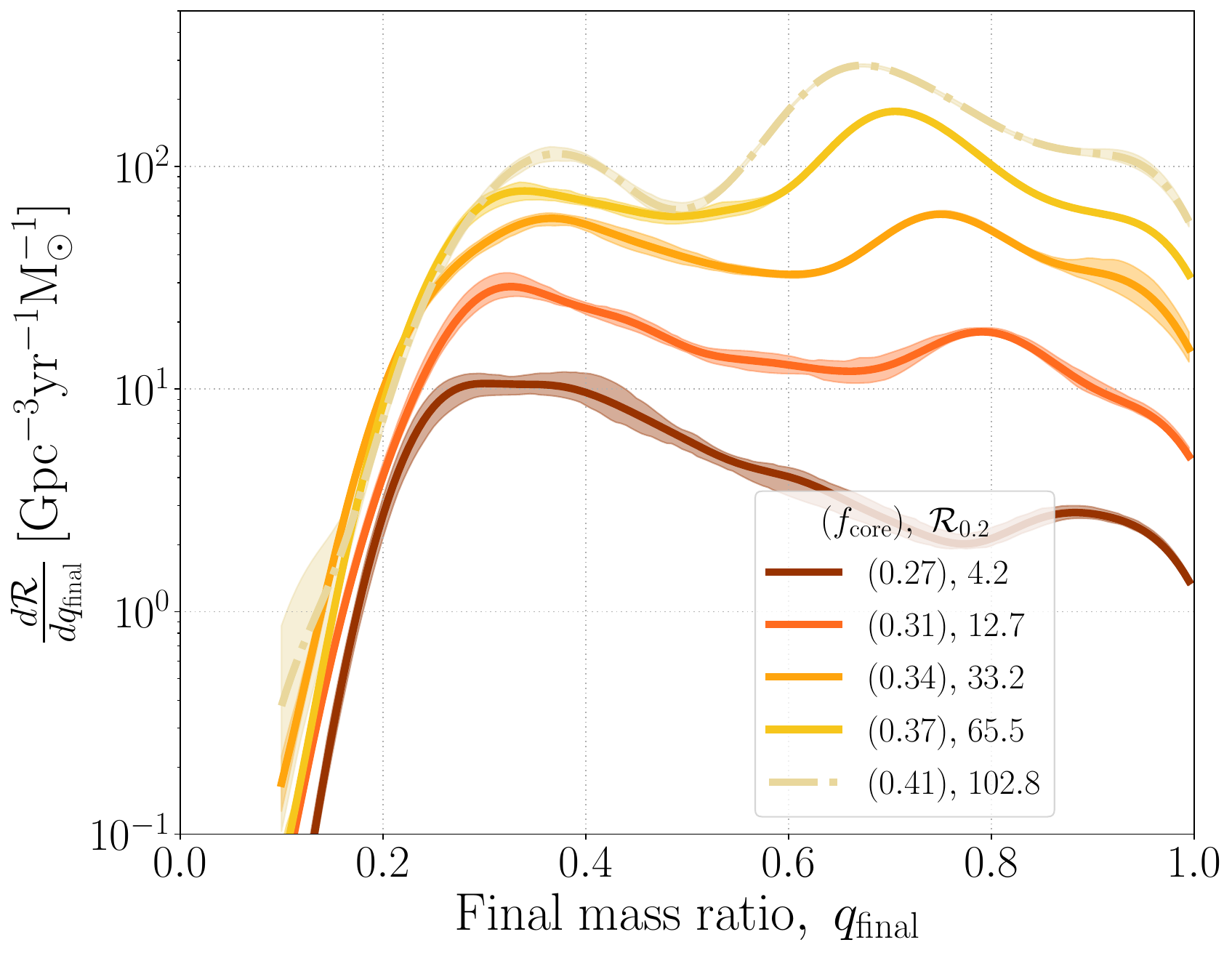}
\caption{The distributions for the chirp mass $M_{\rm Chirp}$, and the final mass ratio $\qfinal = \Mless/\Mmm$ for merging BBH and BHNS. Colours and symbols are the same as the right hand panels of Figure \ref{fig: validation and mass dists}. This shows that the less massive components can form NS masses, often closing any gap between the most massive NS and the least massive BH.
 \label{fig: other dist}}
\end{figure}

In the left column of Figure \ref{fig: other dist} we show the mass distributions for the chirp masses, $M_{\rm Chirp}$, for merging BBH and BHNS from the stable mass transfer channel.
This shows that the less massive components can form masses low enough for NS formation for most variations. Only for the more extreme assumptions of $\zetaeff=3.5$ and $\betaacc=1.0$ does a significant gap remain between the lowest chirp mass and the upper boundary for NS formation (set to $2.5\Msun$ in this work). For almost all variations explored, the distribution of component masses (individual BH and NS masses) does not display an empty `gap' between the most massive NS and the least massive BH.

In the right column of Figure \ref{fig: other dist} we show the final mass ratio $\qfinal = \Mless/\Mmm$.
The mass ratio distributions are all rather flat but display a slight bi-modality with a first peak around $\qfinal \sim 0.35$ and a second peak around $\qfinal\sim 0.75$. This bimodality disappears for $\betaacc = 0.75$ and $\betaacc = 1.0$ because for these mass transfer efficiencies the lower values of $\qfinal$ are excluded.
Similarly, $\zetaeff=3.5$ does not produce any $\qfinal$ near one.
For all physics variations, the mass ratio distribution drops off steeply below  $\qfinal \approx 0.2$, i.e., the stable mass transfer channel is very inefficient at creating the most extreme mass ratio systems.

\section{Overview of rates \label{app: rates} }
{In Tables \ref{tab: rates1} and \ref{tab: rates2} we split \Rlocal, as shown in Figures \ref{fig: validation and mass dists}, \ref{fig: dMsn} and \ref{fig: gamma} into the individual contributions from the BBH and BHNS merger rate. }

\bibliography{main}{}
\bibliographystyle{aasjournal}

\end{document}